\documentclass[lettersize,journal]{IEEEtran}
\usepackage{amsmath,amsfonts}

\usepackage[ruled, linesnumbered, commentsnumbered, longend]{algorithm2e}
\usepackage{graphicx}
\usepackage[table]{xcolor} 
\usepackage{cite}

\usepackage{tikz}
\usepackage{forest}
\usetikzlibrary{trees,positioning,shapes,shadows,arrows.meta}

\usepackage{orcidlink}
\usepackage{multirow}

\usepackage{pdflscape}

\begin{document}

\title{Healthcare Policy Compliance: A Blockchain Smart Contract-Based Approach}

\author{\IEEEauthorblockN{Md Al Amin, Hemanth Tummala, Seshamalini Mohan, and Indrajit Ray}\\
\IEEEauthorblockA{ 
Computer Science Department, Colorado State University, Fort Collins, Colorado,  USA\\
 E-mail: \{Alamin, Hemanth.Tummala, Sesh, Indrajit.Ray\}@colostate.edu
 }
 }

\maketitle

\begin{abstract}
This paper addresses the critical challenge of ensuring healthcare policy compliance in the context of Electronic Health Records (EHRs). Despite stringent regulations like HIPAA, significant gaps in policy compliance often remain undetected until a data breach occurs. To bridge this gap, we propose a novel blockchain-powered, smart contract-based access control model. This model is specifically designed to enforce patient-provider agreements (PPAs) and other relevant policies, thereby ensuring both policy compliance and provenance. Our approach integrates components of informed consent into PPAs, employing blockchain smart contracts to automate and secure policy enforcement. The authorization module utilizes these contracts to make informed access decisions, recording all actions in a transparent, immutable blockchain ledger. This system not only ensures that policies are rigorously applied but also maintains a verifiable record of all actions taken, thus facilitating an easy audit and proving compliance. We implement this model in a private Ethereum blockchain setup, focusing on maintaining the integrity and lineage of policies and ensuring that audit trails are accurately and securely recorded. The Proof of Compliance (PoC) consensus mechanism enables decentralized, independent auditor nodes to verify compliance status based on the audit trails recorded. Experimental evaluation demonstrates the effectiveness of the proposed model in a simulated healthcare environment. The results show that our approach not only strengthens policy compliance and provenance but also enhances the transparency and accountability of the entire process. In summary, this paper presents a comprehensive, blockchain-based solution to a longstanding problem in healthcare data management, offering a robust framework for ensuring policy compliance and provenance through smart contracts and blockchain technology.
\end{abstract}


\section{Introduction} \label{sec:introdduction}

\IEEEPARstart{E}{lectronic} health records (EHRs) have emerged as a cornerstone in modernizing healthcare, offering numerous benefits that enhance efficiency and quality of care. These systems facilitate immediate and remote access to patient data, a critical feature that streamlines the medical care decision-making process. By transitioning from paper-based systems, EHRs significantly reduce errors commonly associated with manual record-keeping, thereby enhancing patient safety and care quality \cite{silow2012using,highfill2019hospitals,jindal2018electronic}. One of the key advantages of EHRs is their ability to promote interoperability across different healthcare platforms. This interconnectedness allows for the seamless sharing of patient data among various healthcare providers, leading to improved continuity of care and a more cohesive healthcare experience. Additionally, the shift to EHRs results in notable cost savings by eliminating the expenses related to paper records and optimizing resources \cite{highfill2019hospitals}.

Furthermore, EHR systems play a vital role in strengthening healthcare services. They enhance clinical cooperation and increase the accuracy of diagnostics. The constant availability of up-to-date patient medical records is crucial for maximizing treatment productivity and ensuring timely and accurate medical interventions \cite{king2014clinical, rani2022survey, cherif2022electronic}. The trend of storing patient information electronically on local databases or cloud servers underscores the healthcare industry's commitment to making patient care more efficient and precise \cite{al2021model}. EHRs represent a significant leap forward in healthcare technology. They streamline administrative processes, improve care coordination, and contribute to better clinical outcomes. As healthcare organizations continue to adopt and refine EHR systems, they pave the way for a more efficient, accurate, and patient-centered approach to healthcare service delivery.

The transition to EHRs has marked a significant advancement in healthcare technology, improving the efficiency and quality of patient care. However, this digital transformation also brings forth complex information security and privacy challenges, which are critical to address for maintaining patient trust and compliance with regulatory standards. Data security and privacy violations are increasingly seen across the healthcare sector. Many of these can be attributed to this industry's widespread use of smartphones, internet-connected devices, sensors, wearable devices, mobile-based health applications, and other IT-dependent services. Cybersecurity risks, potential breaches, and the need for stringent access controls raise significant questions. Additionally, interoperability issues, human factors such as user authentication, and the imperative of legal compliance further compound the challenges associated with implementing and maintaining EHR systems \cite{mathai2017electronic,bayer2015new}.

Laws, policies, and regulations are pivotal in addressing EHR challenges and safeguarding healthcare data security and patient privacy. In various global regions, diverse privacy standards, including the General Data Protection Regulation (GDPR) in Europe, the Health Insurance Portability and Accountability Act (HIPAA) in the United States (US), and My Health Record (MHR) in Australia, have been established to protect patient privacy and personal data \cite{shah2020secondary}. The HIPAA policy standards and guidelines help mitigate the risks associated with EHR and promote trust between patients and healthcare providers.HIPAA requires healthcare organizations to implement technical, administrative, and physical safeguards to secure EHRs \cite{mbonihankuye2019healthcare}. These safeguards include access controls, encryption, authentication measures, and regular security assessments. By enforcing these safeguards, HIPAA helps prevent unauthorized access, data breaches, and identity theft. Furthermore, HIPAA mandates the implementation of privacy policies and procedures to govern the use and disclosure of patient information. It grants patients certain rights, such as the right to access and amend their medical records, and requires healthcare providers to obtain patient consent for certain uses and disclosures of their information.

HIPAA also establishes non-compliance penalties, incentivizing healthcare organizations to prioritize data security and privacy. These penalties can range from fines to criminal charges, depending on the violation severity. The violation can also lead to reputational damage, eroding trust from clients and the public. Organizations may face exclusion from federal programs, financial strain due to legal costs, and increased scrutiny.  The violation of HIPPA can also lead to provider confusion, increased documentation time, alert fatigue, and potential patient safety issues \cite{berg2020resident}. According to the Office of Civil Rights Data study, since October 2009, massive security breaches may have affected more than half of the population in the USA. At least 173 million medical records were breached due to the policy non-compliance. The study showed that the hackers recruited healthcare insiders with access to valuable data `using online ads and social media.

According to the Information Security Media Group, 75\% of US healthcare organizations reported at least one security breach affecting less than 500 individuals, and 21\% reported an incident affecting more than 500 individuals. The Healthcare Information and Management Systems Society found that 68 percent of surveyed healthcare organizations in the US experienced a significant security incident. These incidents were reported to be caused by insider threats (53.7\%) and external threats (63.6\% of healthcare organizations). It is important to note that these reported incidents may not capture all security breaches, as some incidents go undetected or are underreported. Security breaches in healthcare can be costly, with cases of breaches in healthcare data costing hospitals between US\$250000 to US\$2.5M in settlement payments \cite{keshta2021security}.

The landscape of health insurance, designed to safeguard individuals and families against the financial burdens of medical expenses, is now facing a growing challenge from those seeking to exploit vulnerabilities for personal gain due to non-compliance with health care policy. In 2022, 431 individuals were convicted of healthcare fraud, constituting 8.4 percent of all offenses related to theft, property destruction, and fraud. This marks a 1.4 percent rise in healthcare fraud offenders compared to the fiscal year 2018 \cite{usscHealthCare}.  The National Health Care Anti-Fraud Association (NHCAA) estimates that the monetary damages resulting from healthcare fraud reach tens of billions of dollars annually. A statistical analysis suggests it accounts for about 3 percent of healthcare expenditures. However, certain government and law enforcement agencies posit the losses to be as high as 10 percent of the annual health outlay, potentially exceeding 300 billion dollars \cite{NHCAA}. Health insurance fraud has financial repercussions, leading to increased costs for insurers and potentially higher premiums for policyholders. It can compromise the quality of care by encouraging unnecessary medical procedures and overusing services. Providers involved in fraudulent activities may face reputational damage and legal consequences. Overall, health insurance fraud erodes public trust in the healthcare system and diverts resources from critical healthcare initiatives.

\begin{figure}[hbt]
  \centering
  \includegraphics[scale=0.75]{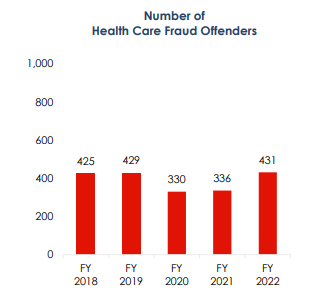}
  \vspace{-1em}
  \caption{Number of Health care Insurance Fraud \cite{usscHealthCare}}
  \label{fig: Number of Health care Insurance Fraud }
\end{figure}

\begin{table}
\centering
\caption{OCR HHS - Compliance Complain} \label{table:ocr-hhs-compliance-complain}
\vspace{-1em}
\scriptsize
\begin{tabular}{|c| c| c| c|c|} 
\hline
Year &  Complains & Compliance Reviews & Technical Assistance & Total Cases   \\
\hline
 2018 & 25089 & 438 & 7243 & 32770 \\ 
 \hline 
 2019 & 29853 & 338 & 9060 & 39251 \\
\hline
 2020 & 26530 & 566 & 5193 & 32289 \\
\hline
 2021 & 26420 & 573 & 4244 & 31237 \\
\hline
\end{tabular}
\end{table}

Table \ref{table:ocr-hhs-compliance-complain} shows the number of compliance complaints received by the U.S. Department of Health and Human Services (HHS) Office for Civil Rights (OCR). The major reasons for the complaints are \textit{(i) impermissible uses and disclosures of PHI; (ii) lack of safeguards of PHI; (iii) lack of patient access to their PHI; (iv) lack of administrative safeguards of electronic PHI;} and \textit{(v) use or disclosure of more than the minimum necessary PHI. }

While advancements in security and privacy technology are essential for enhanced protection of patient data from such incidents, substantial evidence indicates that improper adoption, implementation, and enforcement of policies contribute significantly to unauthorized access—without a legitimate "need to know"—to Electronic Health Record (EHR) data \cite{raghupathi2023analyzing}. Whether intentional or unintentional, users are often granted access privileges they should not possess. Policies are frequently not adhered to accurately, and there are delays in checking or implementing access control rules. Instances have been observed where identical roles and privileges are assigned universally to all employees. Additionally, individual patient-level policies are often not rigorously enforced. Auditing and monitoring gaps are also prevalent, typically occurring only in response to serious complaints or legal mandates.
 
Sarkar et al. \cite{sarkar2020influence} present an alternative perspective on security and privacy breaches within the healthcare industry. They explore the existence of distinct professional subcultures within healthcare organizations. Through a qualitative study, the authors identify factors that inadvertently lead these subculture groups to violate information security policies. For instance, in cases where a doctor, positioned at the apex of the subculture hierarchy, seeks access to information beyond their authorized scope, lower-ranking employees within the subculture may not intervene to prevent it.

\begin{figure}[hbt]
  \centering
  \includegraphics[scale=0.25]{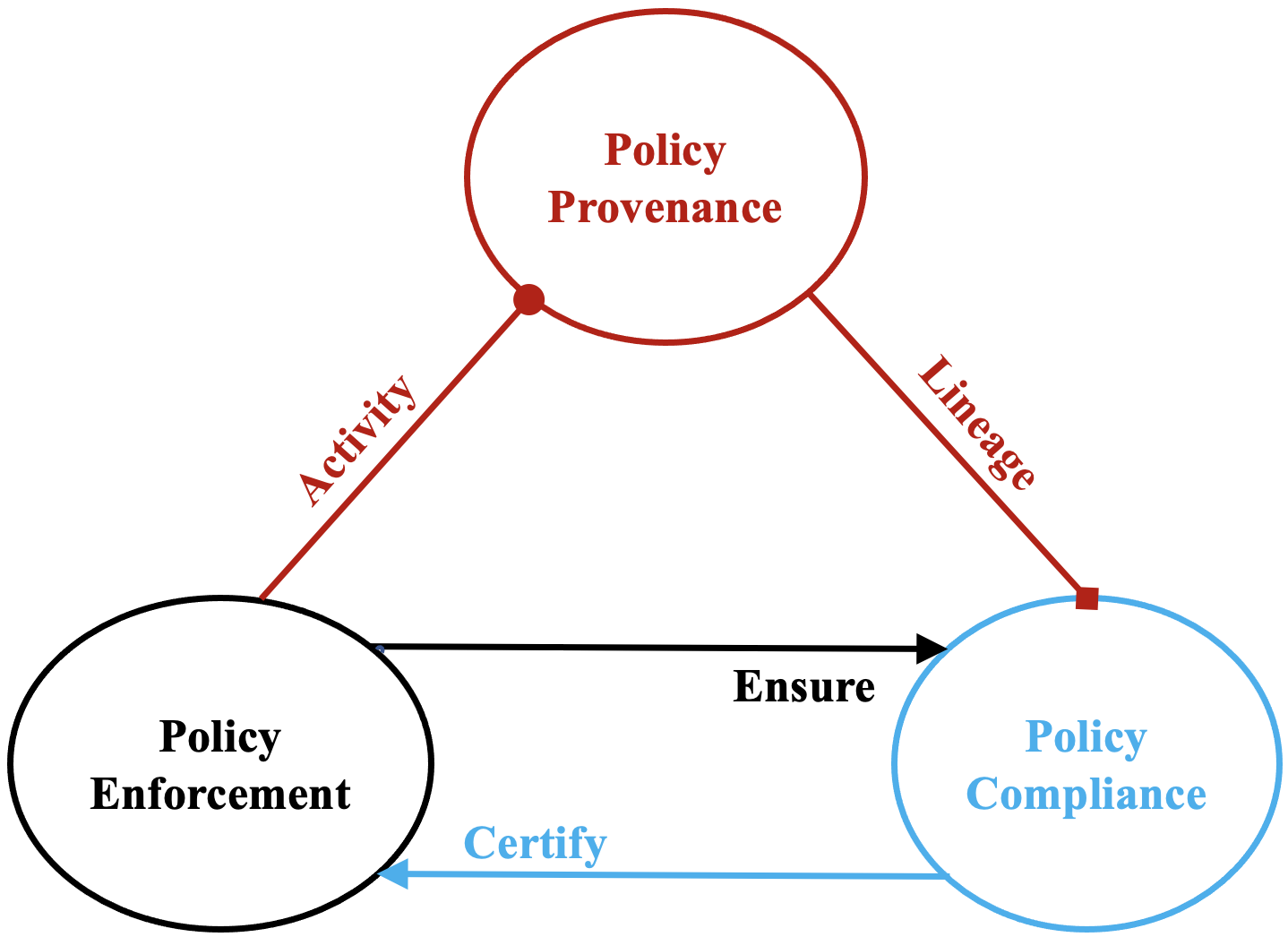}
  \vspace{-.8em}
  \caption{Policy Enforcement, Compliance, and Provenance .}
  \label{fig:policy-enforcement-compliance-provenance}
\end{figure}

While ensuring the enforcement of system security policies is crucial, it is equally vital to establish provenance to validate adherence to these policies. Figure \ref{fig:policy-enforcement-compliance-provenance} illustrates the interplay among enforcement, compliance, and provenance. Effective policy enforcement safeguards healthcare data against unauthorized access and misuse. When policies are adequately enforced, achieving policy compliance becomes possible, as compliance necessitates aligning all actions with the relevant policies. However, this compliance alone lacks quantifiable measurement or validation. To assess policy compliance accurately, maintaining the integrity of policy enforcement activities is indispensable. Enforcement integrity ensures that events are accurately recorded as they unfold. An independent auditor conducts a policy audit to verify the compliance status of the policy. Provenance, in turn, offers a chronological record of policy enforcement activities as they transpire.

Another important factor contributing to policy violations is insufficient enforcement mechanisms due to different sets of policy bodies. Healthcare is often subject to a complex web of regulations and guidelines established by different organizations at the national and international levels. This multiplicity of policy bodies can create challenges and potential conflicts for healthcare providers and professionals \cite{garpenby2022patient,fowler2021implement}. Analysis of data from the United States Department of Health and Human Services, covering data breaches recorded from January 2015 to December 2020, revealed that a significant portion of the compromised data resulted from insufficient communication and training for healthcare professionals. Inadequate communication and training contribute to policy violations in healthcare by leaving staff unaware of existing policies or changes, leading to unintentional non-compliance. Insufficient training exacerbates challenges in understanding complex policies, increasing the risk of improper implementation. Additionally, communication gaps between departments, limited feedback mechanisms, and cultural barriers further hinder effective policy adherence among health professionals \cite{yeo2022human}.

This paper proposes a smart contract-based policy enforcement architecture to accurately enforce deployed authorization, obligation/regulatory, operational policies, and patient-provider agreements (PPAs) during data access decision-making. The approach uses blockchain smart contracts to implement access control features and data security guidelines.  Moreover, we propose a blockchain-based integrity storage system for EHRs, policies, access control model features, audit trails of deployed policies, and the integrity of subjects', objects', and environments' attributes. Since blockchain keeps immutable records, it can keep track of transactions in real-time and identify any unauthorized changes. In addition, the consensus process in a blockchain network guarantees that smart contracts work as intended without user interference.

The proposed architecture enforces applicable policies and PPAs by keeping audit trails linked to enforced policies and enforcing access based on smart contracts. Since the blockchain network stores all user policies and event logs, it provides provenance services. For any transaction, smart contracts will automatically notify the appropriate users. Also, once smart contracts are fully deployed and functional, the conditions and mechanisms written into the code can't be changed.

This paper is an extension of the earlier works \cite{al2023informed, al2023blockchain}. Our major contributions are as follows:
\begin{itemize}
    \item Critical analysis of HIPAA security and privacy and organizational policies to integrate with the proposed framework to ensure policy compliance through proper policy enforcement, preserving policy lineage and enforcement activities as audit trails, and timely compliance checking through auditing the audit trails.
    \item Private or enterprise Ethereum blockchain-based provenance mechanism for storing enforcement activities as audit trails. It guarantees that any audit trail modification can be detected, which is one of the crucial requirements for ensuring and certifying policy compliance through independent audits.
    \item Proof of Compliance (PoC) consensus mechanism to verify the activities' compliance status for the proposed private/enterprise blockchain-based audit trails. A set of distributed, decentralized, and independent auditor nodes performs the compliance checking for the given audit trails. They determine compliance and non-compliance status for every access that happens in the system.
    \item Implementation of the graphical user interfaces (GUI) for the proposed framework with patient-provider agreement (PPA) and informed consent. Through informed consent enforcement, patient privacy policies are integrated and enforced to ensure healthcare policy compliance.
\end{itemize}

The remainder of the paper is organized as follows: Section \ref{sec:related-works} contains some related works. Healthcare policy requirements are discussed in Section \ref{sec:healthcare-policy-compliance-requirements}. Section \ref{sec:proposed-approach-overview} discusses the proposed approach overview or summary for policy enforcement, provenance, and compliance. In Section \ref{sec:policy-enforcement}, the necessary figures, components, and interactions using blockchain and smart contracts are shown for how informed consent can be added to the patient-provider agreement (PPA) and the contact-based enforcement mechanism. A private or enterprise blockchain-based policy provenance mechanism is discussed in Section \ref{sec:policy-provenance} for storing policy enforcement activities as audit trails. The integrity verification process for audit trails is also given. Section \ref{sec:policy-compliance} discusses the requirements and application of a blockchain consensus mechanism called Proof of Compliance (PoC) for the proposed framework to verify the compliance status as compliance and non-compliance for health workers. A sequence diagram for policy enforcement, provenance, and compliance mechanisms is given in Section \ref{sec:achieving-provenance-compliance}. Section \ref{sec:implementation} contains the implementation conceptual designs. The paper is concluded with a brief discussion in Section \ref{sec:conclusion}. Finally, Section \ref{sec:future-directions}  discusses some future research directions to complete and extend the proposed compliance frameworks.

\section{Related Works} \label{sec:related-works}

In the burgeoning landscape of electronic health records (EHR) management, various researchers have proposed innovative blockchain-based frameworks to tackle challenges such as secure storage, scalability, and interoperability. 

Shahnaz et al \cite{shahnaz2019using} proposed a blockchain-based framework for EHR and provided secure storage of electronic records with granular access rules for users. It also addresses the scalability problem of blockchain by utilizing off-chain storage. This paper exclusively concentrates on ensuring secure information storage and does not delve into implementing information consent or other privacy standards mandated by the government.

Mayer et al \cite{mayer2020electronic} analyzed that Blockchain technology can provide secure and tamper-resistant storage of medical records, ensuring data integrity and authenticity, and also suggest that the blockchain directory model and the chain structure of blockchain can support the continuous growth of medical records. The blockchain should implement OpenEHR, HL7 FHIR, HIPAA, GDPR, IHE, ISO, SNOMED, DICOM, HIE, and PII standards to facilitate interoperability and ensure the uniformity of healthcare information for all stakeholders. 

Wang et al  \cite{wang2018secure} developed a combined attribute-based/identity-based encryption and signature blockchain mechanism to minimize the utilization of different cryptographic systems for different security requirements in the EHR. This system ensures the integrity and traceability of medical data. This work only focuses on the encryption of the signatures and identity of the patient and doesn’t focus on safeguarding other patient information like health history, insurance company details, etc.  

Azaria et al \cite{azaria2016medrec} designed a  blockchain prototype to handle the permission and data access management system called MedRec [4]. MedRec provides patients with a comprehensive and unalterable record of their medical information, facilitating easy retrieval from their healthcare providers and treatment locations. The framework supports interoperability, improves the data quality and the quantity of the medical records, and reduces the fragmentation and latency while accessing the medical records. Our system also manages permission and data access using blockchain, similar to the MedRec architecture. It ensures transparency in patient record access by seeking patient consent when sharing the records with team members.

Amin et al\cite{al2023blockchain} introduced a blockchain-based contract-oriented access control model aimed at enforcing agreements between patients and providers (PPAs) and other healthcare policies to address issues in policy compliance that may result in unauthorized access [5]. Their proposed architecture utilizes blockchain to store policies, access control measures, audit trails, and the integrity of Electronic Health Records (EHRs) attributes. Integrating patient-provider agreements with the access control model enhances transparency and accountability. Our model extends this framework by incorporating policy compliance and an informed consent mechanism.

Albalwy et al.\cite{albalwy2021blockchain} introduced ConsentChain, a blockchain-based dynamic consent management architecture on the Ethereum platform, facilitating clinical genomic data exchange. This system utilizes smart contracts to model patient actions, data creators, and data requesters for granting or revoking data-sharing permissions. However, the focus of the work is primarily on genomic data sharing with specific stakeholders, such as clinicians and researchers, leaving out considerations for the distinct consent management requirements in clinical treatment scenarios. Recognizing the distinct requirements in clinical treatment processes, our proposal suggests a consent management framework to address complex permission assignments for diverse roles like treatment team members, insurance agents, external doctors, and pharmacists.

Tith et al. \cite{tith2020patient} propose an electronic consent management model using Hyperledger Fabric blockchain and purpose-based access control. Patient records, consents, and metadata are stored on the blockchain and shared among participating organizations, with a Chaincode managing the business logic for consent. While the initial model is designed for data sharing and donation, the authors highlight the limitations of the Hyperledger network in terms of restricted participation and limited public trust. To address this, we advocate for the adoption of the private Ethereum blockchain for its broader participant inclusion, enhanced transparency through a public consensus mechanism, and widespread use of smart contracts in various projects.

Haque et al. \cite{haque2021towards} introduced an architecture for a COVID vaccination passport (VacciFi) that adheres to GDPR by storing vaccination data in off-chain storage. They utilize a permissioned blockchain to enhance the ability of participating entities to monitor activities. This study offers valuable insights into the design requirements of our proposed system.

Piao et al. \cite{piao2021data} suggested a data-sharing scheme for GDPR compliance based on blockchain. The goal is to encourage adherence to regulations and offer a platform for user-service provider interaction to ensure secure data sharing. This study primarily emphasizes user authentication and private data sharing, with less emphasis on provenance.

The existing study suggested the potential effectiveness of blockchain in storing compliance-related information for provenance; most of them primarily focus on regulatory policy compliance and do not address individual policies within patient consent and organizational access control policies. We are deploying the patient policy smart contract in the blockchain framework to uphold the integrity of the patient's records.

\section{Healthcare Policy Compliance Requirements} \label{sec:healthcare-policy-compliance-requirements}

Healthcare policies vary significantly across countries, influenced by political systems, economic conditions, cultural values, and historical developments. In India, the regulatory framework includes the \textit{Personal Data Protection Bill} and the \textit{Information Technology Act}. European countries adhere to the \textit{General Data Protection Regulatio}n; Australia utilizes \textit{My Health Record (MHR)}; the USA follows HIPAA; and Canada abides by the \textit{Personal Information Protection and Electronic Documents Act (PIPEDA)}. The US government also follows certain federal laws in addition to international laws to maintain the integrity and confidentiality of the data. The \textit{Emergency Medical Treatment and Labor Act} is a federal law enacted in 1986 that requires hospitals to provide emergency medical treatment to individuals regardless of their ability to pay or their insurance status. It prohibits patient dumping, which is the refusal of care or transfer of patients with unstable medical conditions. 

\textit{Children's Health Insurance Program (CHIP)} is a joint federal and state program that provides health coverage to children in low-income families who do not qualify for Medicaid but cannot afford private insurance. In 2016, the \textit{21st Century Cures Act} was passed as a federal law aimed at expediting advancements in health research. It addresses both privacy concerns, and the law also prohibits the practice of information blocking, wherein organizations engage in activities hindering or preventing access to electronic health information. Additionally, the 21st Century Cures Act introduces provisions enabling the compassionate sharing of mental health and substance abuse treatment details with family members and caregivers. Violations of this prohibition can result in fines of up to 1 million dollars per instance. 

Apart from international and federal policies, health care must even abide by state laws to address the privacy and security of medical records, like \textit{California Confidentiality of Medical Information Act (CMIA)}, \textit{Colorado Medical Records Privacy Act (CMRPA)}, Arizona \textit{Health Information Exchange (HIE)}, etc. The local and city governments also play a supportive role in healthcare through public health department rules, community health initiatives, etc., which overarch the regulatory framework established by federal and state laws. The HIPAA policy also allows the organization to set up its policy and rules to ensure data security. The Facility Access Control and Workstation Security rules of the physical safeguard security policy are established and managed by organizational laws. We are implementing HIPAA and HITECH policies within our framework to ensure data protection.

\subsection{HIPAA Overview}
\begin{figure}[tb]
      \begin{center}
        \includegraphics[scale=0.5]{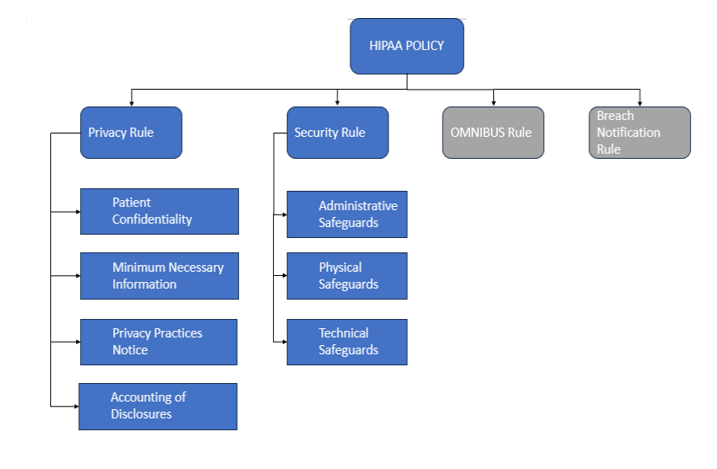}
        \caption{Types of HIPAA Rules.}
        \label{fig:hipaa-rules-classification}
      \end{center}
    \end{figure}

HIPAA, or the Health Insurance Portability and Accountability Act, is a U.S. federal law enacted in 1996 that establishes standards and safeguards for the protection of sensitive patient health information, known as Protected Health Information (PHI)\cite{(OCR)_2023}. Ensuring adherence to the policies outlined by the government in HIPAA is a crucial aspect when constructing any information security framework. This is particularly important because patients value the confidentiality of their healthcare records. Patients who do not trust the healthcare framework may withhold essential information from healthcare providers. The HIPAA policy is divided into four rules as shown in Fig. \ref{fig:hipaa-rules-classification}: \textit{(i) privacy rule, (ii) security rule, (iii) omnibus rule,} and \textit{(iv) breach notification rule.}

The HIPAA Privacy Rule outlines the standards for protecting individuals' medical records and other personal health information, known as protected health information (PHI), held by covered entities and their business associates. The Privacy Rule sets forth rules and procedures that covered entities must follow to ensure the confidentiality and privacy of PHI. It establishes the rights of individuals over their health information, including the right to access their records, request corrections, and control the disclosure of their health information.

The HIPAA Security Rule is a set of regulations established under the Health Insurance Portability and Accountability Act (HIPAA) to protect electronic protected health information (ePHI). It outlines standards and safeguards for covered entities, such as healthcare providers and health plans, that must be followed to secure ePHI. The rule encompasses administrative, physical, and technical safeguards, requiring entities to implement policies, procedures, and technologies to safeguard electronic health information's confidentiality, integrity, and availability. 

The HIPAA Omnibus Rule introduced significant modifications to strengthen the HIPAA policy. It expanded liability by making business associates directly accountable for compliance with certain HIPAA provisions, particularly the Security Rule. The rule heightened breach notification requirements, specifying what constitutes a reportable breach and establishing notification procedures. Additionally, it incorporated amendments related to the Genetic Information Nondiscrimination Act (GINA), further safeguarding genetic information from misuse in health plans. 

The HIPAA Breach Notification Rule mandates that covered entities and their business associates notify affected individuals, the U.S. Department of Health and Human Services (HHS), and, in certain cases, the media in case of a breach of protected health information (PHI). A breach is the unauthorized acquisition, access, use, or disclosure of PHI that compromises its security or privacy. Covered entities must conduct a risk assessment to determine the probability of compromise, and if the risk is low, breach notification requirements may be waived. Non-compliance with breach notification rules can result in financial penalties.

The US government also launched the HITECH Act (\textit{Health Information Technology for Economic and Clinical Health Act}) which introduced several policies and provisions aimed at promoting the adoption and meaningful use of health information technology, particularly electronic health records (EHRs). HITECH made significant changes to HIPAA by amending its rules and strengthening its enforcement mechanisms. It introduced new requirements and increased penalties for non-compliance. HITECH expanded the liability of business associates by holding them directly accountable for compliance with certain HIPAA provisions. Business associates are now subject to many of the same rules and penalties as covered entities. HITECH also significantly increased the penalties for HIPAA violations, providing a tiered structure based on the level of negligence. The maximum annual penalty for a single violation increased substantially. HITECH also brought forth additional mandates regarding data breach notifications. According to the HITECH Breach Notification Rule, in the event of a data breach, HIPAA-covered entities are obligated to inform the individuals impacted by the breach. Furthermore, notification must be provided to both the Secretary of Health and Human Services and the media if the breach affects more than 500 individuals.

\subsection{HIPAA Regulated Organizations}  
\begin{figure*}[tb]
    \centering
    
    \tikzset{
        basic/.style  = {draw, text width=3cm, align=center, font=\sffamily, rectangle},
        root/.style   = {basic, rounded corners=2pt, thin, align=center, fill=green!30},
        onode/.style = {basic, thin, rounded corners=2pt, align=center, fill=green!60,text width=3cm,},
        tnode/.style = {basic, thin, align=left, fill=pink!60, text width=15em, align=center},
        xnode/.style = {basic, thin, rounded corners=2pt, align=center, fill=blue!20,text width=5cm,},
        wnode/.style = {basic, thin, align=left, fill=pink!10!blue!80!red!10, text width=6.5em},
        edge from parent/.style={draw=black, edge from parent fork right}
        }
        
    \begin{forest} for tree={
        grow=east,
        growth parent anchor=west,
        parent anchor=east,
        child anchor=west,
        edge path={\noexpand\path[\forestoption{edge},->, >={latex}] 
            (!u.parent anchor) -- +(10pt,0pt) |-  (.child anchor) 
            \forestoption{edge label};}
    }
    [HIPAA Regulated Organizations, basic,  l sep=10mm,
        [Business Associates, xnode,  l sep=10mm, [Lawyers\, , tnode] [Financial Organizations\, , tnode] [Cloud Service Providers\, , tnode] [Software Providers\, , tnode]]
        [Covered Entity, xnode,  l sep=10mm, [Health Clearing House\, , tnode] [Health Plans\, , tnode] [Healthcare Providers\, , tnode]
        ]]
    \end{forest}

    \caption{HIPAA Regulated Organizations.} \label{fig:hipaa-regulated-organizations}
\end{figure*}
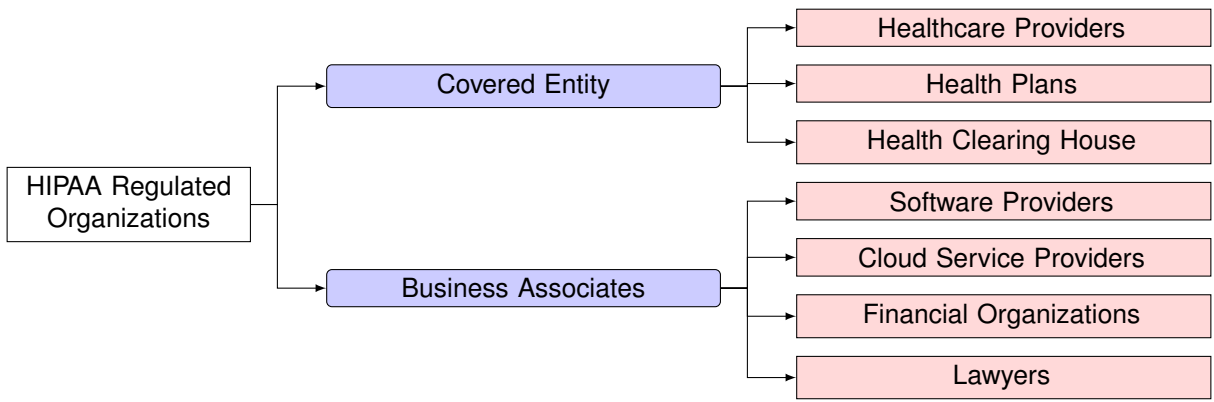

The regulations outlined in the HIPAA Rules pertain to both covered entities and business associates. Entities, organizations, and agencies falling within the scope of a covered entity as defined by HIPAA are obligated to adhere to the Rules. These regulations mandate safeguarding the privacy and security of health information while also granting individuals specific rights regarding their health data. In cases where a covered entity enlists the assistance of a business associate for healthcare-related activities, a formal contract or arrangement must be in place. Fig. \ref{fig:hipaa-regulated-organizations} shows the HIPAA-regulated organizations.

The HIPAA new regulation typically mandates that covered entities and their business associates establish contracts, termed a business associate agreement (BAA) to ensure proper protection of protected health information. These contracts also serve to define and restrict, as necessary, the acceptable uses and disclosures of protected health information by the business associate. According to the updated regulations, business associates are now directly accountable under HIPAA, and they face enforcement actions in the same manner as covered entities.

A formal agreement between a covered entity and a business associate must include an outline of the allowed and mandatory uses and disclosures of protected health information by the business associate, and stipulate that the business associate will not utilize or disclose the information beyond what is permitted by the contract or as mandated by law, and mandate the implementation of adequate safeguards by the business associate to prevent unauthorized use or disclosure, including adherence to the requirements of the HIPAA Security Rule concerning electronic protected health information, etc.

\subsection{HIPAA Rules Incorporation In Framework}
The proposed framework incorporates the privacy and security rules of the HIPAA/HITECH policy to protect patient information, adheres to the rules of the regulated organization, and doesn't implement the omnibus and breach notification rules as shown in Fig. \ref{fig:hipaa-rules-classification}. The framework adheres to the following privacy and security guidelines:

\begin{itemize}
    \item The patient grants consent for the treatment team members to access their information. The patient also controls the extent of information shared with specific team members and the type of access, such as read, write, and update permissions. As a result, the privacy and HIPAA policies related to authorization and disclosure accounting comply.
    \item The framework enables patients to grant access rights to their emergency contacts, allowing them to access information. It also allows restricted access to specific information, ensuring compliance with the confidential communications policy.
    \item Due to frequent updates in HIPAA policies, it is crucial for those managing patient information to stay informed about these changes. The healthcare prototype manages workforce training and security following HIPAA policies by implementing an expiration date for HIPAA training. Authorities cannot access patient health records once their HIPAA training expires, and access is only reinstated after they undergo the required training renewal.
    \item Password authentication and information consent mechanisms are employed within the framework to enforce HIPAA policies related to access and audit control. Access to information is restricted to authorized entities, and the system generates regular audit logs, notifying relevant authorities.
    \item The workstation security physical safeguard security policy is enforced as the framework is a licensed application and will be installed on secured workstations. Restricting the application license to a select number of fully authenticated systems installed on workstations helps minimize the risk of data breaches.
\end{itemize}

The developed framework offers the benefit of enhancing transparency for patients by informing them about who is accessing their information. This instills confidence in patients, assuring them that their data is secure and aligns with most privacy and security policies outlined in HIPAA.


\section{Proposed Approach Overview} \label{sec:proposed-approach-overview}

To ensure healthcare policy compliance, it is essential to enforce the right set of policies through the right enforcement mechanism. It is also equally important to ensure provenance regarding applicable policy lineage and policy enforcement activities or audit trails. Finally, independent auditors must verify the audit trails to determine compliance and non-compliance status. If the right set of policies is enforced timely and all provenance is preserved, then the organizations or systems comply with the appropriate policies and best practices. The relationship between policy enforcement, provenance, and compliance is depicted in Fig. \ref{fig:policy-enforcement-compliance-provenance}.

The main idea is to integrate components of informed consent into the patient-provider agreement. Then create and deploy smart contracts for informed consent components in the blockchain network. The authorization module calls the corresponding smart contract for an access request to enforce informed consent. The request specifies which subject wants to perform which operation on what objects under what constraints or conditions. Once the smart contract is called and the authorization module decides, the corresponding event information is recorded as logs in the blockchain network.

To accurately evaluate compliance, we must keep a comprehensive record of policy history and the exact execution of enforcement actions. This history includes all the policies used by the authorization system to make decisions. Ensuring the accuracy of recorded events is vital, as these reflect actual enforcement actions. The records provide a clear sequence of enforcement steps, safeguarding against unauthorized modifications of audit trails or inappropriate access to healthcare data. A private blockchain-based provenance system is proposed where all audit trails are stored. Section \ref{sec:policy-provenance} includes the detailed mechanism for maintaining provenance.

To ensure that systems adhere to set policies, a consensus mechanism called Proof of Compliance (PoC) is proposed. PoC involves a network of independent, decentralized, and independent auditor nodes. These nodes are responsible for auditing and verifying whether the system's operations and access permissions comply with the required policies from available provenances. All these enforcement checks are recorded in an 'audit blockchain', providing a transparent and reliable history of compliance. By using the Proof of Compliance mechanism, organizations can more effectively tackle and reduce compliance-related challenges. Section \ref{sec:policy-compliance} discusses all the required components for policy compliance activities.

The auto-triggering feature of smart contracts ensures that relevant event or activity data is captured and recorded reliably, without omitting any essential details. This data is then stored securely on the blockchain network, a platform known for its decentralized, immutable, and tamper-proof nature. Such a robust system guarantees that the original agreements made between patients and providers are upheld. It also ensures the integrity of event logs, maintaining them exactly as they were created and recorded, free from any unauthorized alterations

\section{Policy Enforcement}  \label{sec:policy-enforcement}

Policy enforcement refers to the implementation of rules, procedures, and safeguards to ensure compliance with the regulations and requirements outlined in HIPAA. This includes enforcing policies related to the security and privacy of protected health information (PHI). Enforcement mechanisms are essential for holding covered entities, business associates, and other entities accountable for safeguarding PHI and complying with the HIPAA rules.

\subsection{Patient-Provider Agreement (PPA)}
The patient-provider agreement aims to determine who is responsible for what in treatment. The goal is to improve outcomes, lower risks, and educate patients better. A multi-center study \cite{pergolizzi2017multicentre} evaluated the utility of the PPA, how readily patients understood it, its ability to educate patients in an unbiased way about treatment, and the feasibility of incorporating a PPA in clinical practice. Both patients and doctors believe this PPA helped them decide on a course of treatment and was fair in laying out the treatment's risks and benefits. Most patients reported the PPA to be "somewhat helpful" or "very helpful" in deciding on a course of treatment and "easy to understand."  A PPA, also known as a contract, differs from organization to organization. Healthcare organizations adjust what they need from patients and what they expect from them to match those needs, treatments, and responsibilities. This is done based on the nature and needs of treatment and services. Also, the components and representation of the PPA depend on the hospital or clinic. Examples include general hospitals, emergency rooms, urgent care or walk-in clinics, dental care, cancer treatment, physiotherapy, etc.

\begin{figure*}[tb]
    \centering
    \includegraphics[scale=0.5]{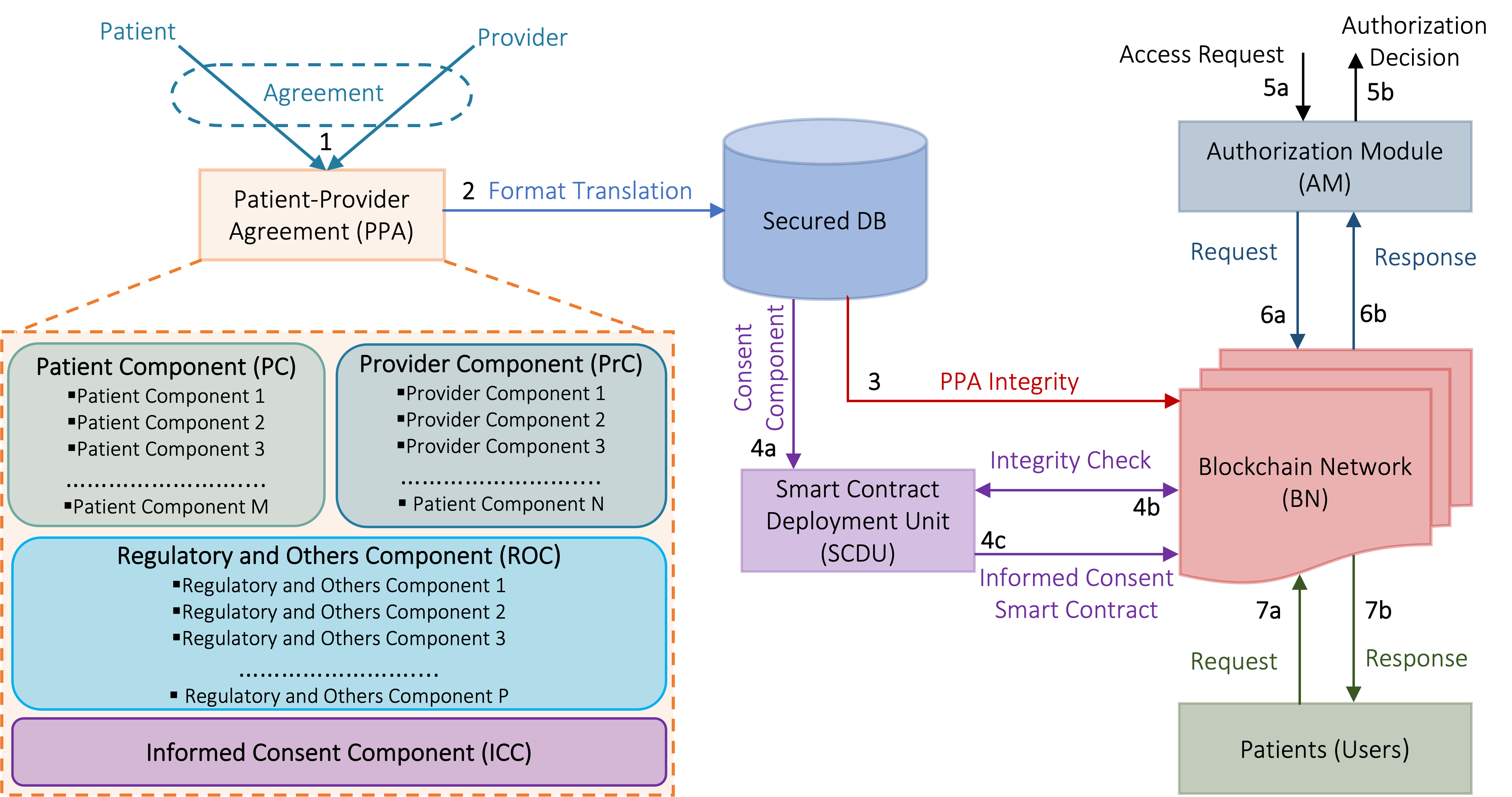}
    \caption{Patient Provider Agreement (PPA) and  Informed Consent Management Framework \cite{al2023informed}.}
    \label{fig:patient-provider-agreement}
\end{figure*}

The patient-provider agreement is depicted in Figure \ref{fig:patient-provider-agreement} with the necessary components. The Patient-Provider Agreement formally is composed of four tuples: \[ PPA = (PC, PrC,  ROC, ICC) \] satisfying the following requirements:
\begin{itemize}
    \item[(i)] $PC$ is a finite set of patient components containing the patient's \textit{personal information, contact information, mailing information, pharmacy information, billing and insurance information, emergency contact}, and \textit{others}. The patient is responsible for providing and maintaining valid information for these components.
    
    \item[(i)]  $PrC$ is a finite set of provider components, including the treatment team, anonymous data sharing for research, prescription, and others. Treatment team members for a patient include \textit{doctors, nurses, support staff, lab technicians, and billing officers}. During the treatment period for a patient, everything from treatment to insurance coverage and billing is considered.
 
    \item[(iii)] $ROC$ is a finite set of regulatory and other components. It has applicable security and privacy policies to comply with the local government, regulatory agencies (HIPAA, GDPR), federal government, and foreign government requirements if necessary.

    \item[(iv)] $ICC$ is a finite set of informed consent components. It indicates that the patient has given permission to access medical data.
\end{itemize}
  
Algorithm \ref{alg:patient-provider-agreement} shows the step-by-step instructions for creating a PPA with $PC$, $PrC$, $ROC$, and $ICC$. A patient-provider agreement is formed when a patient visits a hospital. The terms and conditions of the contract make it invalid after a certain period. There may be several contracts for a single patient. Several patient-provider agreements must be created and properly documented to deliver healthcare services. Managing many contracts involves various things, such as contract creation, development, testing, updating, etc. If the requests contain contracts, the authorization module must consider those with other required policies when making access decisions. From Figure \ref{fig:patient-provider-agreement}, it is seen that the proposed model stores the integrity of a PPA on the blockchain network to ensure the detection of any modification, intentionally or unintentionally.

\RestyleAlgo{ruled}
\SetKwComment{Comment}{/* }{ */}

\begin{algorithm}[tb]
     \footnotesize
    \SetKwInOut{KwData}{Input}
    \SetKwInOut{KwResult}{Result}
    \DontPrintSemicolon

\caption{Patient-Provider Agreement (PPA)}\label{alg:patient-provider-agreement}
\KwData{(i) $PC$, (ii) $PrC$, (iii) $ROC$, (iv) $ICC$, (v) $\mathbb{R}_{PPA}$, (vi) $\mathbb{BN}_{SC}$}
                \textcolor{blue}{\Comment*[r]{$\mathbb{R}_{PPA}$: secured PPA repository, $BN_{SC}$: blockchain network smart contract}}
\KwResult{A formal $PPA$}

\textbf{Input Parameters Initialization} \vfill 
  $PPA_i \gets  \{ PC_i, PrC_i, ROC_i, ICC_i\}$ for patient identity $i$ \vfill
    \quad \textit{(i)} $PC \gets \{ \wp_1, \wp_2, \wp_3,\wp_4, \wp_5,........\wp_M \}$\; \vfill
    \quad \textit{(ii)} $PrC \gets \{ \delta_1, \delta_2, \delta_3,\delta_4, \delta_5,........\delta_N \}$\; \vfill
    \quad \textit{(iii)} $ROC \gets \{ \Re_1, \Re_2, \Re_3,\Re_4, \Re_5,........\Re_P \}$\; \vfill
     \quad \textit{(iv)} $ICC \gets \{ 	\ell_1, \ell_2, \ell_3,\ell_4, \ell_5,........\ell_R \}$\; \vfill
 
 \textbf{PPA Components Integrity Calculation} \vfill        \textcolor{blue}{\Comment*[r]{$\mathbb{H}(\partial)$ calculates hash of $\partial$}}
    \quad \textit{(a)} $\mathbb{H}_{PC} \gets \{ \wp_1, \wp_2, \wp_3,\wp_4, \wp_5,........\wp_M \}$\; \vfill
    \quad \textit{(b)} $\mathbb{H}_{PrC} \gets \{ \delta_1, \delta_2, \delta_3,\delta_4, \delta_5,........\delta_N \}$\;  \vfill
    \quad \textit{(c)} $\mathbb{H}_{ROC} \gets \{ \Re_1, \Re_2, \Re_3,\Re_4, \Re_5,........\Re_P \}$\;\vfill
    \quad \textit{(d)} $\mathbb{H}_{ICC} \gets \{ 	\ell_1, \ell_2, \ell_3,\ell_4, \ell_5,........\ell_R \}$\;  \vfill
    \quad \textit{(e)} $\mathbb{H}_{PPA_i} \gets  \mathbb{H}( \mathbb{H}_{PC}, \mathbb{H}_{PrC}, \mathbb{H}_{ROC}, \mathbb{H}_{ICC})$\; \vfill

 \textbf{PPA Finalization} \vfill 
  \eIf{$PPA_i$ is complete}{
        \textcolor{blue}{\Comment*[r]{complete: presence of $PC$, $PrC$, $ROC$, $ICC$}}
            \eIf{$(\mathbb{R}_{PPA} + PPA_i)$ contains no conflicts}{    
                   \textit{(i)} do $\mathbb{R}_{PPA} \gets (\mathbb{R}_{PPA} + PPA_i)$\;
                   \textit{(ii)} add $\mathbb{ID}_{PPA_i}$ to patient profile, $\mathbb{P}_i$\;
                   \textit{(iii)} call $\mathbb{BN}_{SC} (\mathbb{ID}_{PPA_i}, \mathbb{H}_{PPA_i})$\; 
                        \textcolor{blue}{ \Comment*[r]{later PPA integrity verification}}
                   
                   \textbf{\textit{Return: }} Success ($PPA_i$ added to $\mathbb{R}_{PPA}$)\;
            }{
                \textbf{\textit{Error: }} $(\mathbb{R}_{PPA} + PPA_i)$ contains conflicts\;
                            \textcolor{blue}{ \Comment*[r]{ $PPA_i$ revision required to add to $\mathbb{R}_{PPA}$}}
            }
   }{
        \textbf{\textit{Error: }}$PPA_i$ cannot be created (incomplete PPA)\;  
 }
\end{algorithm}

\subsubsection{Informed Consent Components}
\begin{figure*}[tb]
    \centering
    \includegraphics[scale=0.5]{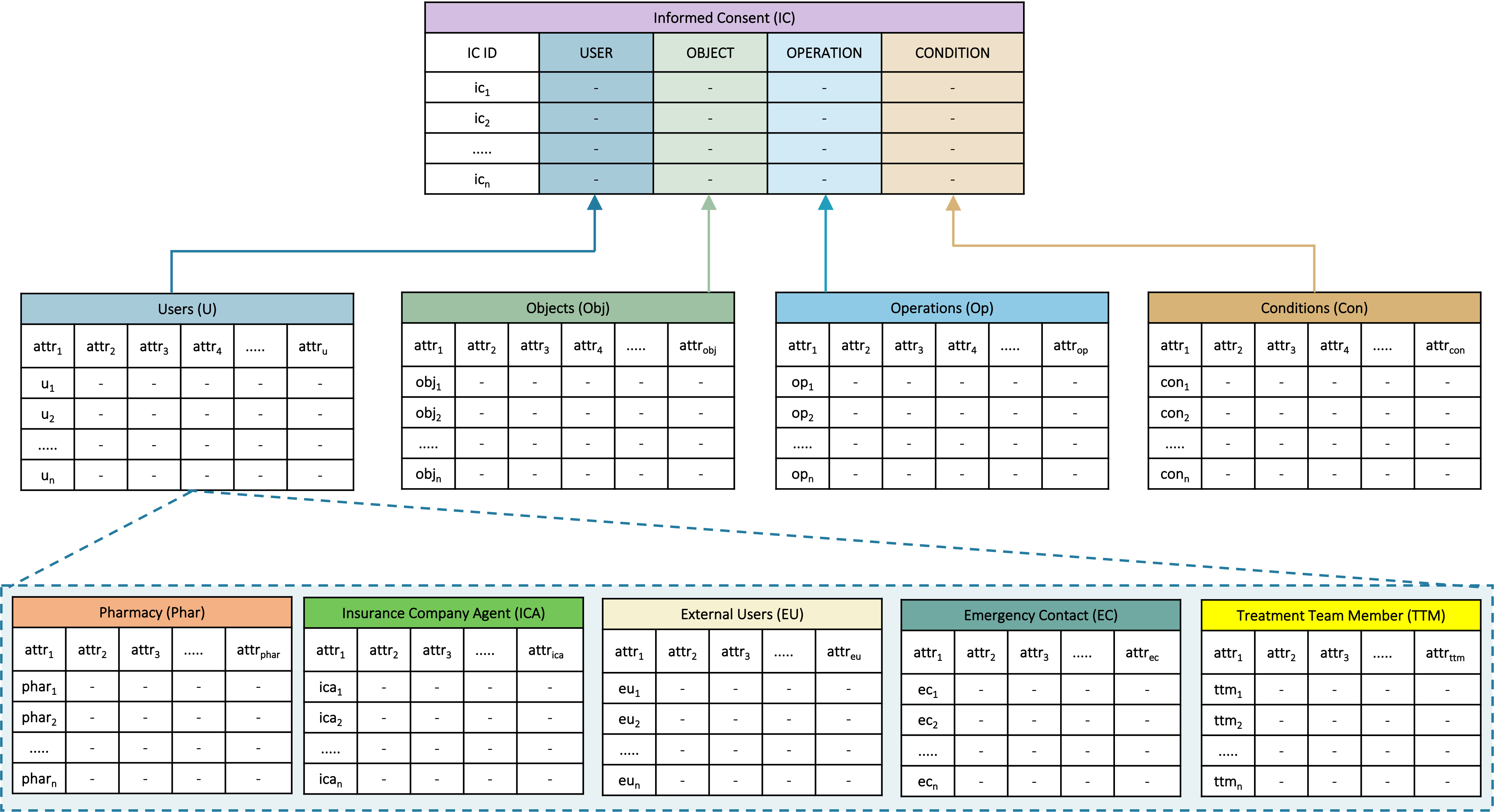}
    \caption{Informed Consent Components \cite{al2023informed}.}
    \label{fig:informed-consent-components}
\end{figure*}

Before giving consent, patients need to know everything about the particular consent. Figure \ref{fig:informed-consent-components} shows the informed consent conceptual framework structure. The Informed Consent formally is composed of four tuples: \[ IC = (U, O,  OP, CON) \]  satisfying the following requirements:
\begin{itemize}
    \item[(i)] $U$ is a finite set of authorized users denoted as $\lbrace u_1, u_2, u_3, .....\rbrace$. The user can perform certain operations on healthcare resources when certain conditions are satisfied.
    
    \item[(i)]  $O$ is a finite set of protected objects otherwise known as protected healthcare resources. A finite set of protected objects $(O)$ denoted as $\lbrace o_1, o_2, o_3, .....\rbrace$.
 
    \item[(iii)] $OP$ is a finite set of operations denoted by  $\lbrace op_1, op_2, op_3, ...\rbrace$. Operations represent the system actions that authorized users can perform on the objects. Examples of operations are read, write, and update.

    \item[(iv)] $CON$ is a finite set of conditions. It indicates the conditions that must be satisfied by the user to perform operations on the protected objects. A finite set of conditions, $CON$, can be denoted as $\lbrace con_1, con_2, con_3, ...\rbrace$.
\end{itemize}

There are many users in the healthcare system. Each user plays a different role and responsibility in performing their job. Treatment team members for a patient include doctors, nurses, support staff, lab technicians, billing officers, the patient's emergency contact person, and other hospital employees assigned by the authority.  Some outsider members are insurance agents, pharmacists or pharmacy technicians, doctors or lab technicians from another hospital. As the treatment period for a patient, everything from treatment to insurance coverage and billing is considered. Informed consent users can be anyone from five groups of people: \textit{(i) treatment team member, (ii) emergency contract, (iii) external users, (iv) insurance company agent,} and \textit{(v) pharmacy}. External users are from different hospitals when a patient is transferred for better treatment if the situation demands it. Usually, external users have temporary access to admitted patients' health records.

The term object refers to an electronic version of a patient's medical history kept on file by the healthcare provider over time. It may include all the administrative and clinical information pertinent to the patient's care under a specific provider, such as demographics, progress notes, issues, medications, vital signs, previous medical history, immunizations, laboratory information, and radiology reports. These objects must be protected from unauthorized users. The main purpose of informed consent permitting by patients to users to perform certain operations.

Those who are qualified to perform necessary operations in the healthcare sector carry out several operations. Some common operations are \textit{view/read}, \textit{add/write}, \textit{update/modify}, \textit{delete}, etc. In the \textit{view} operation, users can only view or read healthcare records or resources if the request is valid and complies with all applicable policies. The state of the data is not changed in this operation, ensuring data integrity. However, it can compromise confidentiality and privacy if the access requested is granted without appropriate credentials. On the other hand, the \textit{write} operation changes the state of the records or healthcare data. If proper policy enforcement is not ensured, it breaks the integrity of the data.

There might be various constraints or conditions under which certain consent can be enforced, rejected, revoked, and others. The conditions can be but are not limited to:

\begin{itemize}
    \item[(i)] \textbf{\textit{Time Constraints:}} In time constraints, any user can access a patient's healthcare data within a certain time. For example, the time condition for consent is regular office hours: 8 am-5 pm. In this case, the request is rejected if any subject wants to access the patient's record beyond this time. The attempt is recorded as an audit trail event.
    
    \item[(ii)] \textbf{\textit{Date Constraints:}} The date constraints limit the calendar date. No access request is granted beyond the intended date.

    \item[(iii)] \textbf{\textit{Day Constraints:}} Day conditions can include work days (Monday-Friday), weekends (Saturday–Sunday), holidays, etc. Based on the day, the subject can access data. Suppose a regular doctor has a duty on workdays. On weekends, no access is given to that doctor. 

    \item[(iv)] \textbf{\textit{Location-based Constraints:}} The location-based condition allows users to access information from a certain location, like a hospital building, inside an emergency room for treating emergency patients, and others.

    \item[(v)] \textbf{\textit{IP-based Constraints:}} IP-based conditions limit healthcare users from accessing resources from certain IPs. Device IPs must be from the known list; otherwise, no access is granted.

    \item[(vi)] \textbf{\textit{Access Frequency Dependent:}} A user can operate for a certain number in access-frequency-dependent conditions. Suppose an external doctor is given five times view permission. Once the doctor reads the patient's specified records five times, the given consent is expired, and access is denied. There is no access without getting new consent.
\end{itemize}

The above mentions list is not fixed for the conditions, but we consider them for this study. There might be other conditions depending on the nature of the treatment, patient characteristics, provider business policy, nature, etc. With sophisticated technology, malicious attackers can spoof the conditions to fool the system into accessing healthcare data and other compromised credentials. Proper layered defense mechanisms must be deployed to ensure that conditions' credentials are accurate, not fabricated or manipulated.

\subsubsection{Informed Consent Smart Contract Generation} \label{subsec:consent-contract-generation}
Once a patient-provider agreement, or PPA, is created and stored in the repository, all informed consent components are deployed as smart contracts. The patient owns all the deployed contracts. The authorization module needs to access these smart contracts to make decisions with other components such as subject attributes, object attributes, operations attributes, environmental attributes, organizational policies, regulatory and other policies, and others required. Algorithm \ref{alg:consent-solidity-contract-deployment} shows the steps to develop and deploy smart contracts for informed consent components. The smart contract deployment unit, SCDU, collects all consent components from PPA and checks integrity to confirm that collected consents are not modified deliberately or inadvertently. In step 3 in Figure \ref{fig:patient-provider-agreement}, PPA integrity as the hash from Algorithm \ref{alg:patient-provider-agreement} ($Hash_{PPA_i}$) is stored in the blockchain network along with PPA id.

To verify PPA integrity, SCDU calls the corresponding smart contract function to retrieve the PPA hash value stored in the network. Any modification of consent components voids consent. If there is no modification, then SCDU creates and deploys smart contract(s) to the blockchain network. Once the contracts are deployed, the contract addresses are added to the patient's profile and hospital systems. The contract address is an identifier for a smart contract in the blockchain network.

\RestyleAlgo{ruled}
\SetKwComment{Comment}{/* }{ */}

\begin{algorithm}[tb]
     \footnotesize
    \SetKwInOut{KwData}{Input}
    \SetKwInOut{KwResult}{Output}
    \DontPrintSemicolon

\caption{Informed Consent Smart Contract}\label{alg:consent-solidity-contract-deployment}

\KwData{Informed Consent Component $ICC$.}
\KwResult{Smart contract contains informed consents elements $ICC$ from  patient-provider agreement $PPA$.}
\textbf{Initialization} \vfill 
\textit{Informed Consent} $IC_i :=  \{ Sub, Op, Obj, Cond\}$ where $Sub, Op, Obj, Cond$ represent one or more individual attribute(s) and $i$ represents patient identity;\vfill
\textit{Subject Attributes} $Sub := \{ SubAttr_1, SubAttr_2,........SubAttr_M \}$ \vfill
\textit{Operation Attributes} $Op := \{ OpAttr_1, OpAttr_2,........OpAttr_N \}$ \vfill
\textit{Object Attributes} $Obj := \{ ObjAttr_1, ObjAttr_2,........ObjAttr_P \}$ \vfill
\textit{Conditions} $Cond := \{ CondAttr_1, CondAttr_2,........CondAttr_R \}$ \vfill

 \textbf{Smart Contract Generation and Deployment} \vfill
  \eIf{$IC_i$ is complete}{
                    \textcolor{blue}{\Comment*[r]{complete means presence of $Sub$, $Op$, $Obj$, and $Cond$ with patient consent}}
        $IC_i$ is added to smart contract\; 
   }{
       \textit{Denied: } smart contract for $IC_i$ can not be created and deployed\;  \textcolor{blue}{\Comment*[r]{incomplete informed consent component}}
 }
\end{algorithm}

\subsection{Contract-Based Access Control}
\begin{figure*}[tb]
    \centering
    \includegraphics[scale=0.5]{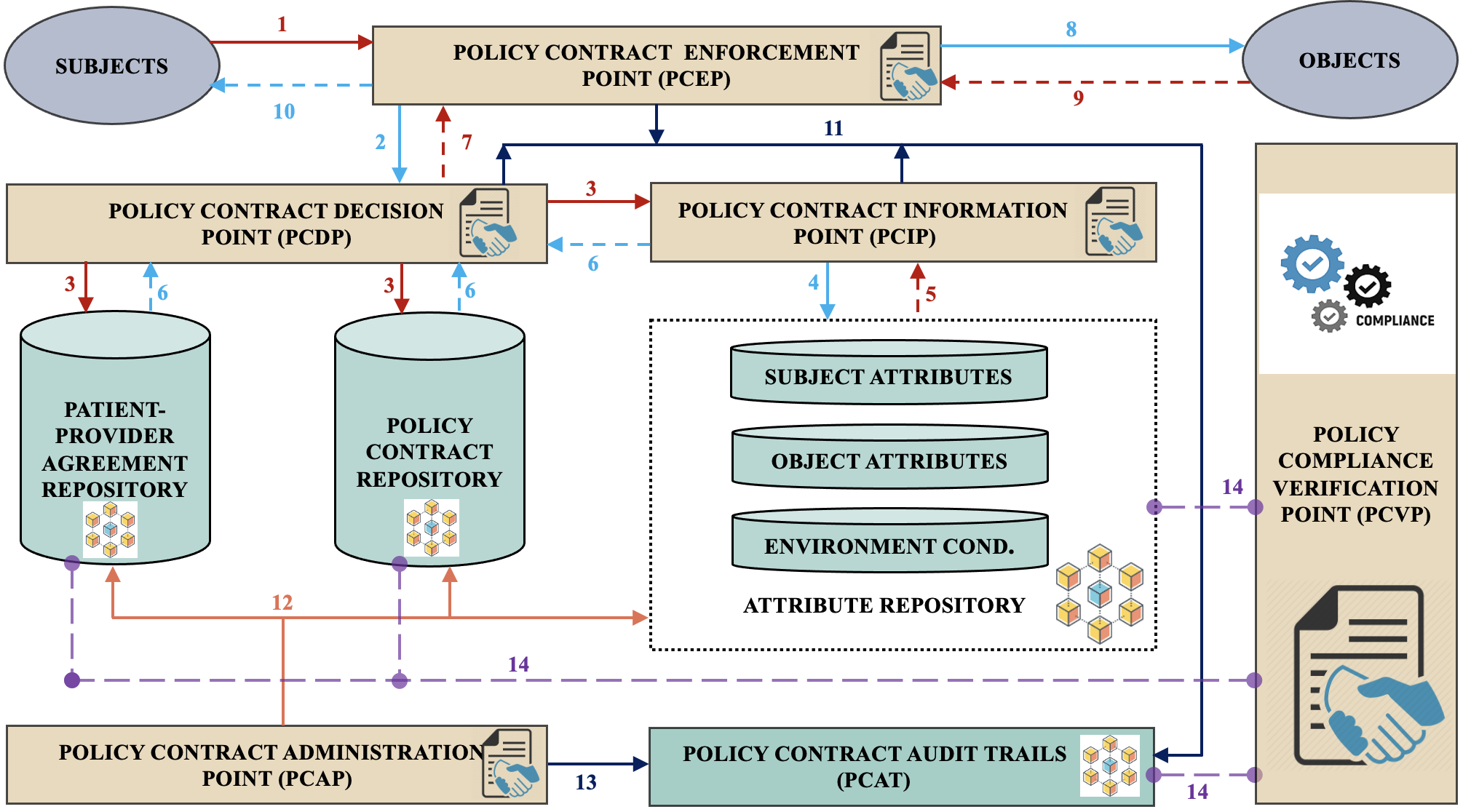}
    \vspace{-1em}
    \caption{Contract-Based Access Control Model \cite{al2023blockchain}.}
    \label{fig:contract-based-access-control-model}
\end{figure*}

We propose an extended version of the Attribute-Based Access Control Model \cite{hu2015attribute} to integrate the Patient-Provider Agreement (\textit{PPA}) for authorizations with other applicable policies and attributes. The proposed model is called the Contract-based Access Control Model. Integrating PPA into the access control model improves transparency and accountability and facilitates compliance monitoring. Figure \ref{fig:contract-based-access-control-model} contains proposed model components. The solid line with the arrow indicates request/command and the dotted line with the arrow means response/feedback. The dotted line with circles (purple colored) shows the request/command and response/feedback. Their interactions and descriptions are discussed in the following.

\textbf{Subject (SB):} A subject is a human user or non-person entity (NPE), such as a device that issues access requests to perform operations on objects. Subjects are assigned one or more attributes. A total $m$ number authorized subjects can be represented as $\{sb_1, sb_2, sb_3, sb_4,.....sb_m\}$.

\textbf{Subject Attributes (SA):} Subject attributes of a subject, such as a name, date of birth, home address, training record, and job function, may, individually or combined, comprise a unique identity that distinguishes that user from all others. A finite number, total $n$, of subject attributes defined as $\{sa_1, sa_2, sa_3, sa_4,.....sa_n\}$.

\textbf{Object (OB):} An object can be a resource or requested entity and anything upon which a subject may operate, including data, applications, services, devices, and networks. A finite number, total $p$, of objects to be protected can be written as $\{ob_1, ob_2, ob_3, ob_4,.....ob_p\}$.

\textbf{Object Attributes (OA):}  An object's attributes help to describe and identify it. Attributes include the object name, creator, creation time, and others. A $q$ number of object attributes can be expressed as $\{oa_1, oa_2, oa_3, oa_4,.....oa_q\}$.

\textbf{Operation (OP):} An operation is an action that can be requested by any subject for any object. Only a subject can be authorized to perform a requested operation on an object if the policy allows it. A finite, total $r$, of actions to be performed are denoted as $\{op_1, op_2, op_3, op_4,.....op_r\}$.

\textbf{Environment Condition (EC):}  Environmental conditions are dynamic factors, independent of subject and object, that may be used as attributes at decision time. They may include location, time, day of the week, threat level, device ID, user IP address, temperature, etc. A finite, total $s$, of environmental conditions can be expressed as $\{ec_1, ec_2, ec_3, ec_4,.....ec_s\}$. 

\textbf{Policy Contract Administration Point (PCAP):} It provides functionalities for creating, storing, managing, and testing policies, PPAs, and SB, OB, and EC attributes. 

\textbf{Attributes Repository (AR):} It contains all SB, OB, and EC attributes. Before storing, \textit{PCAP} ensures the authenticity and integrity of the attributes' sources and values because the repository acts later as a standard to verify and compare the user's provided attributes to make the authorization decisions.

\textbf{Policy Contract Repository (PCR):}  The \textit{PCR} contains digital policies (DPs) and metapolicies (MPs) of obligatory policies like organizational policies, regulatory agency policies, and access control policies. A policy is the representation of rules that makes it possible to determine if requested access should be allowed, given attributes of the SB, OB, and EC.

\textbf{Patient-Provider Agreement Repository (PPAR):}  It contains all valid contracts made by patients and providers. The \textit{Policy Contract Decision Point or PCDP} must execute \textit{PPAs} related to an access request.

\textbf{Policy Contract Decision Point (PCDP):} It computes access decisions by evaluating the applicable policies from \textit{PCR}, PPAs from \textit{PPAR}, and attributes from \textit{AR}.

\textbf{Policy Contract Information Point (PCIP):} The \textit{PCIP} is the retrieval source of the attributes required for policy evaluation to make authorizations by \textit{PCDP}.

\textbf{Policy Contract Enforcement Point (PCEP):} After making authorizations, the \textit{PCDP} forwards decisions to \textit{PCEP}. The \textit{PCEP} can access protected resources and objects through a resource access point (\textit{RAP}). Multiple \textit{RAP} might exist, but every object is accessible only through a single \textit{RAP}. 

\textbf{Policy Contract Audit Trails (PCAT):} It contains the activities happening in the system and related to policy compliance. All activities other than policy compliance-related are not required for this proposed model. The actions can be reconstructed properly from \textit{PCAT} to recreate the events performed by \textit{PCEP, PCDP, PCIP, PCAP}, and \textit{SB}.

\section{Policy Provenance} \label{sec:policy-provenance}

While enforcing policy for system security is essential, it is equally important to maintain provenance to demonstrate policy compliance. However, by itself, policy compliance cannot be measured or validated. An independent auditor performs a policy audit to certify the policy's compliance status using available provenance data. To measure policy compliance, it is essential to maintain the following:
\begin{itemize}
    \item [-] policy lineage
    \item [-] integrity of policy enforcement activities
\end{itemize}
Policy lineage contains all policies where the authorization module makes authorization decisions based on these policies. Enforcement integrity means the events are recorded as they happened. Provenance provides a lineage of policy enforcement activities as they are executed. No one should be able to modify audit trails to cover activities or unwanted healthcare data access. This section contains the detailed mechanisms of provenance to ensure policy lineage and audit trail integrity.

\subsection{Provenance via Blockchain} \label{sec:provenance-service}
\begin{figure}[ht]
    \centering
    \includegraphics[scale=0.5]{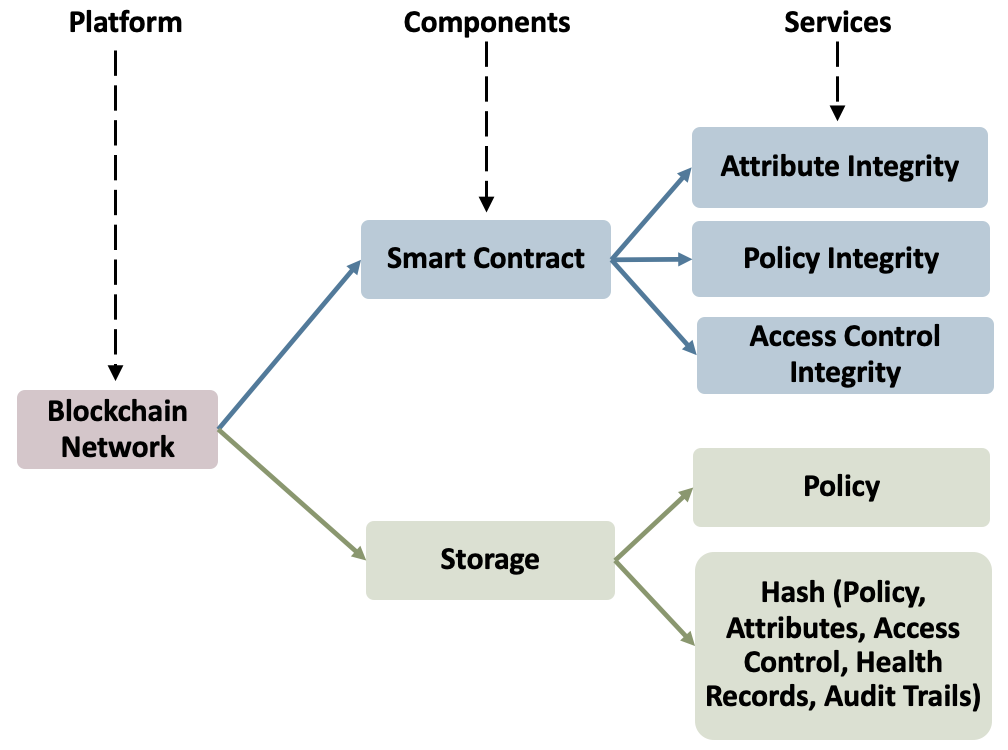}
    \vspace{-1em}
    \caption{Provenance via Blockchain.}
    \label{fig:provenacne-via-blockchain}
\end{figure}

Figure \ref{fig:provenacne-via-blockchain} depicts the provenance services via the blockchain network. The blockchain network acts as a platform that has two major components: \textit{(i) smart contract} and \textit{(ii) storage capacity}. The smart contract provides integrity services for attributes, policies, and access control mechanisms. The blockchain network stores policy, the hash of policy, attributes, access control, health records, audit trails, and others.

\subsection{Policy Enforcement Audit Trails}
\begin{figure}[tb]
    \centering
    \includegraphics[scale=0.5]{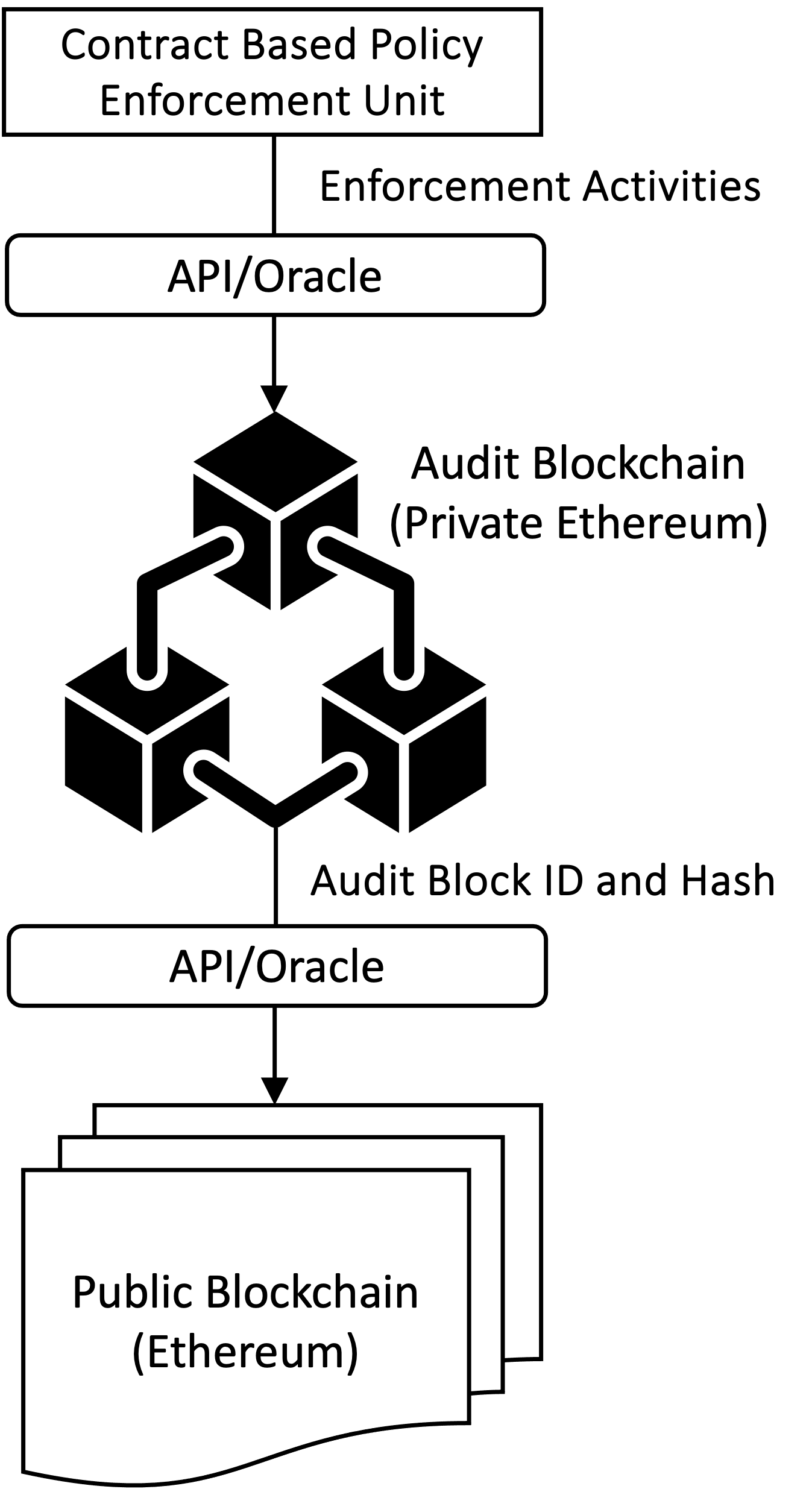}
    \vspace{-1em}
    \caption{Policy Enforcement Audit Trails.}
    \label{fig:policy-enforcement-audit-trails}
\end{figure}

\begin{figure}[tb]
    \begin{center}
    \includegraphics[scale=0.7]{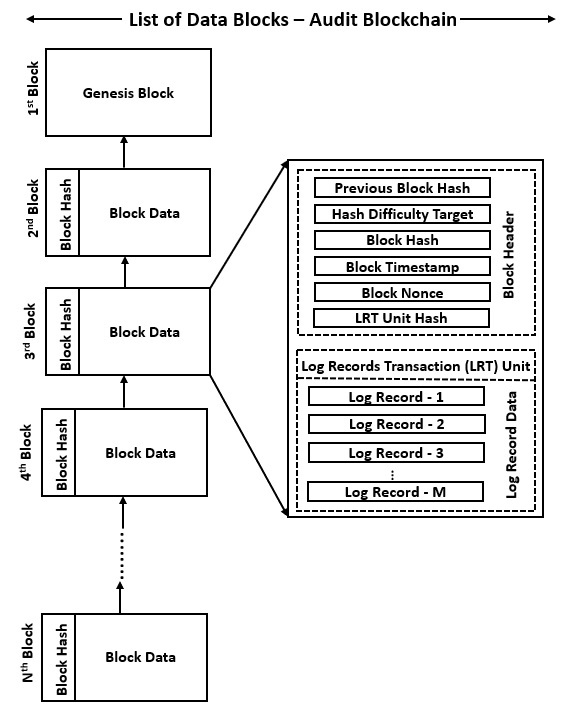}
       \vspace{-1em}
    \caption{Audit Blockchain Block Structure.}
    \label{fig:audit-block-structure}
    \end{center}
\end{figure}

We propose a private blockchain-based audit trail-storing system. All policy enforcement activities are collected and stored on the private blockchain network as audit trails. The private Ethereum blockchain network is used to store audit trails. Fig. \ref{fig:policy-enforcement-audit-trails} shows the structure of policy enforcement audit trails. A private blockchain is a setup where a closed group of users determines the consensus mechanism and other properties like block structure, block size, and block contents. To ensure that audit trails are not intentionally modified, this proposed approach stores block ID and block hash (as integrity) on a public blockchain like Ethereum. Fig. \ref{fig:audit-block-structure} depicts the audit block structure for the audit blockchain. Each block contains a $M$ number of audit trails as transactions, where each log includes which subject performs what operation on which object under what conditions at what time.

An API, or Oracle, is used to interact with a blockchain network from another one. Since smart contracts cannot communicate directly from one blockchain network to another. API/Oracle is a trusted and blind entity that reads data from the source and transfers it to the destination without modifying or revealing data to other entities. The detailed mechanism of the API or Oracle is out of the scope of this paper. We will address the complete technical design and implementation in our future communication.

Storing policy enforcement activities can be done using database management systems. However, we propose a private blockchain (Ethereum private or enterprise setup) instead of database systems. In Section \ref{sec:policy-compliance}, a consensus mechanism called Proof of Compliance (PoC) is proposed to verify the compliance status of enforcement activities. A set of dedicated blockchain nodes perform compliance-checking operations according to the consensus algorithm. Also, the existing smart contract framework can be used to perform various operations to provide services, such as all audit trails for a particular user or an object, and so on. To execute the PoC consensus mechanism, storing audit trails on a blockchain is required. It is very expensive to store data on public blockchain networks. Moreover, systems generate a large amount of data, and audit trails contain sensitive information regarding users' actions. We avoid storing audit trails on the public blockchain.

\subsection{Audit Trail Verification}
\begin{figure}[tb]
    \centering
    \includegraphics[width=0.45\textwidth]{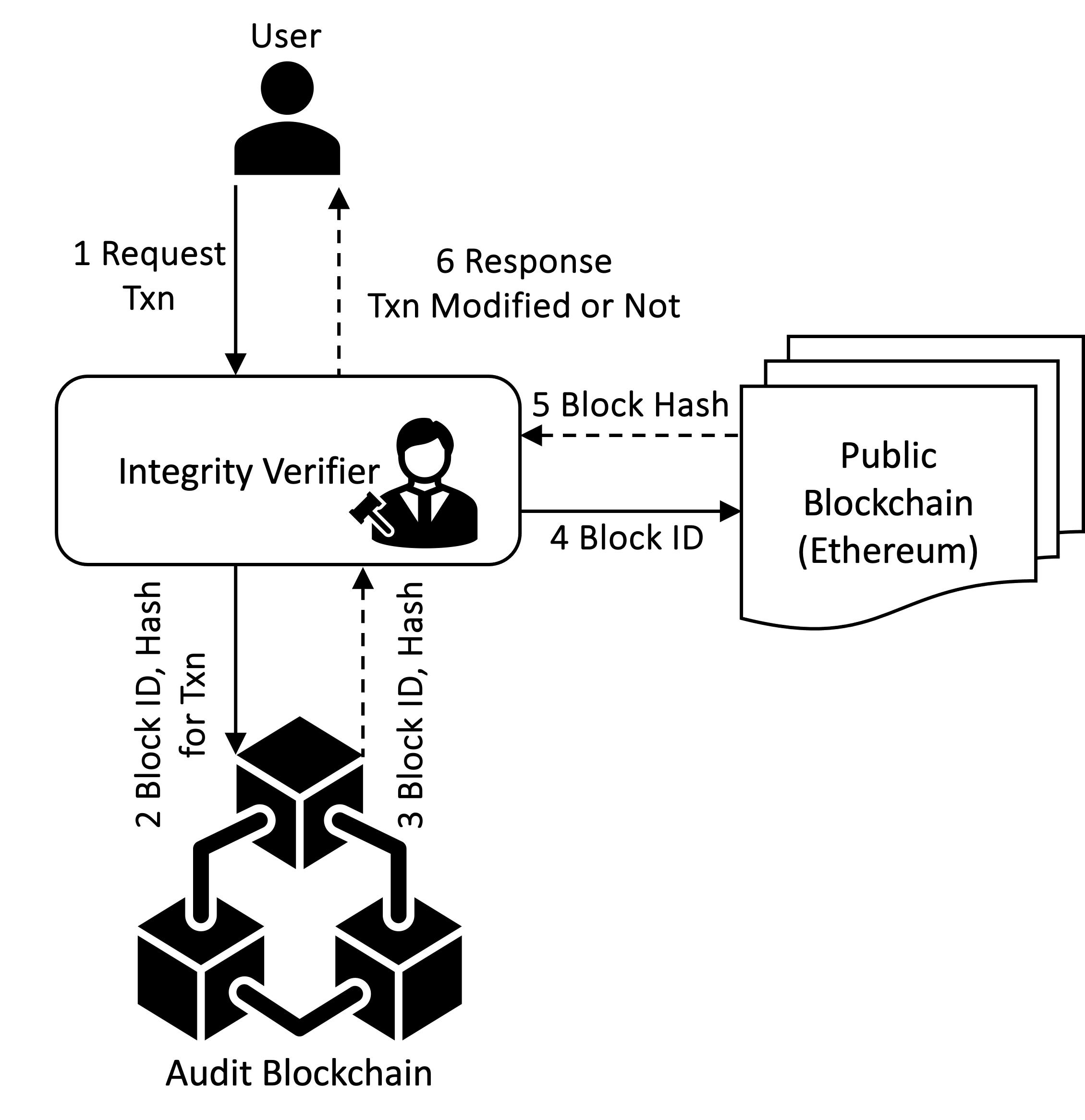}
    \vspace{-1em}
    \caption{Audit Trails Integrity Verification}
    \label{fig:audit-trail-integrity-verification}
\end{figure}

Everyone should not access the audit trails since they contain sensitive information regarding users' executed operations or health data access. Access to audit logs must be controlled and restricted to a certain group of users who have privileges according to the organization and other applicable policies. However, it is necessary to access or check a user's activity data for various purposes. So, there must be a process to verify the integrity of a user's audit trails from the audit blockchain without revealing other users' activity data. We propose a zero-knowledge-based audit trail integrity verification process shown in Fig. \ref{fig:audit-trail-integrity-verification}. A user provides audit trails or activity data and gets modified or not-modified responses from the system. After getting a request from a user, the Integrity Verifier (IV) gets the block ID and integrity from the audit blockchain. Then IV queries the public blockchain to retrieve the block hash for the audit blockchain block ID. If audit block integrity and stored hashes on the public blockchain are matched, then IV returns are not modified or tampered with by the user. Otherwise, the activity data is tampered with in the audit blockchain. Moreover, all blocks in the audit blockchain are added through a consensus mechanism. A single-bit modification invalidates all the blocks in the tampered block. Which indicates any kind of modification. Here, the Integrity Verifier also acts as a blind and trusted entity, such as an API or Oracle. It does not reveal data to other entities and analyzes data for any interest. It also does not modify any data while processing users' requests.

\section{Policy Compliance} \label{sec:policy-compliance}

The healthcare industry is subject to varying degrees of regulatory oversight, and compliance with these regulations is essential for its operation and growth. The specific regulatory landscape can vary by country and region, adding complexity to this industry. Non-compliance or compliance failure causes various business and legal issues, like medical and business license suspension, employee termination, monetary fines, business restrictions within a particular jurisdiction, business reputation loss, patient or client dissatisfaction, etc. These non-compliance cases are mostly found or noted while conducting internal, external, or third-party audits and reviews. To help healthcare organizations detect early non-compliance issues, this section presents a consensus mechanism called Proof of Compliance ($PoC$). Some distributed, decentralized, and independent auditor nodes check the compliance status of any logical operations or accesses that have already been approved, granted, or executed in the system to get to PHI. The $PoC$ auditors work on audit trail provenance data that is securely stored and maintained on the audit blockchain. Section \ref{sec:policy-provenance} contains the detailed proposed mechanism of maintaining a private blockchain-based audit trail provenance system. The Proof of Compliance consensus mechanism helps organizations minimize compliance challenges. Organizations can consider $PoC$ outputs and take further actions to reduce non-compliance cases to avoid compliance issues and business losses.

\subsection{Transaction Structure}
\begin{figure}[tb]
      \begin{center}
        \includegraphics[scale=0.4]{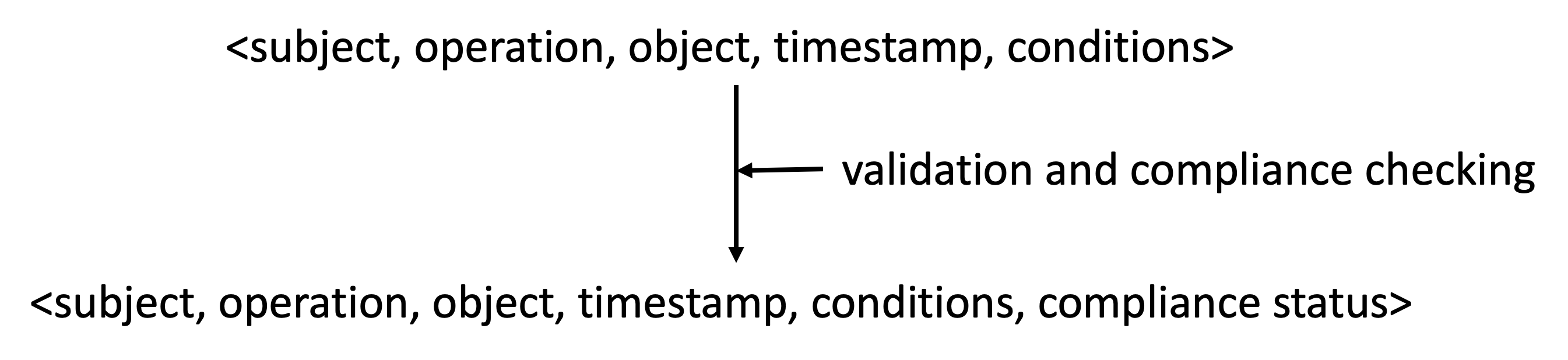}
        \caption{Proof of Consensus (PoC) Consensus Transaction Structure.}
        \label{fig:transaction-structure}
      \end{center}
    \end{figure}

Fig. \ref{fig:transaction-structure} shows the conceptual transaction structure for the proposed consensus mechanism. An individual transaction represents an approved or granted health data access request. Initially, there are five elements in any transaction: \textit{(i) subject, (ii) operation, (iii) object, (iv) conditions, } and \textit{(iv) timestamp.} The system adds one more element, \textit{compliance status}, to the initial transaction after performing a compliance status check.

    \begin{figure}[tb]
    \begin{center}
    \includegraphics[scale=0.45]{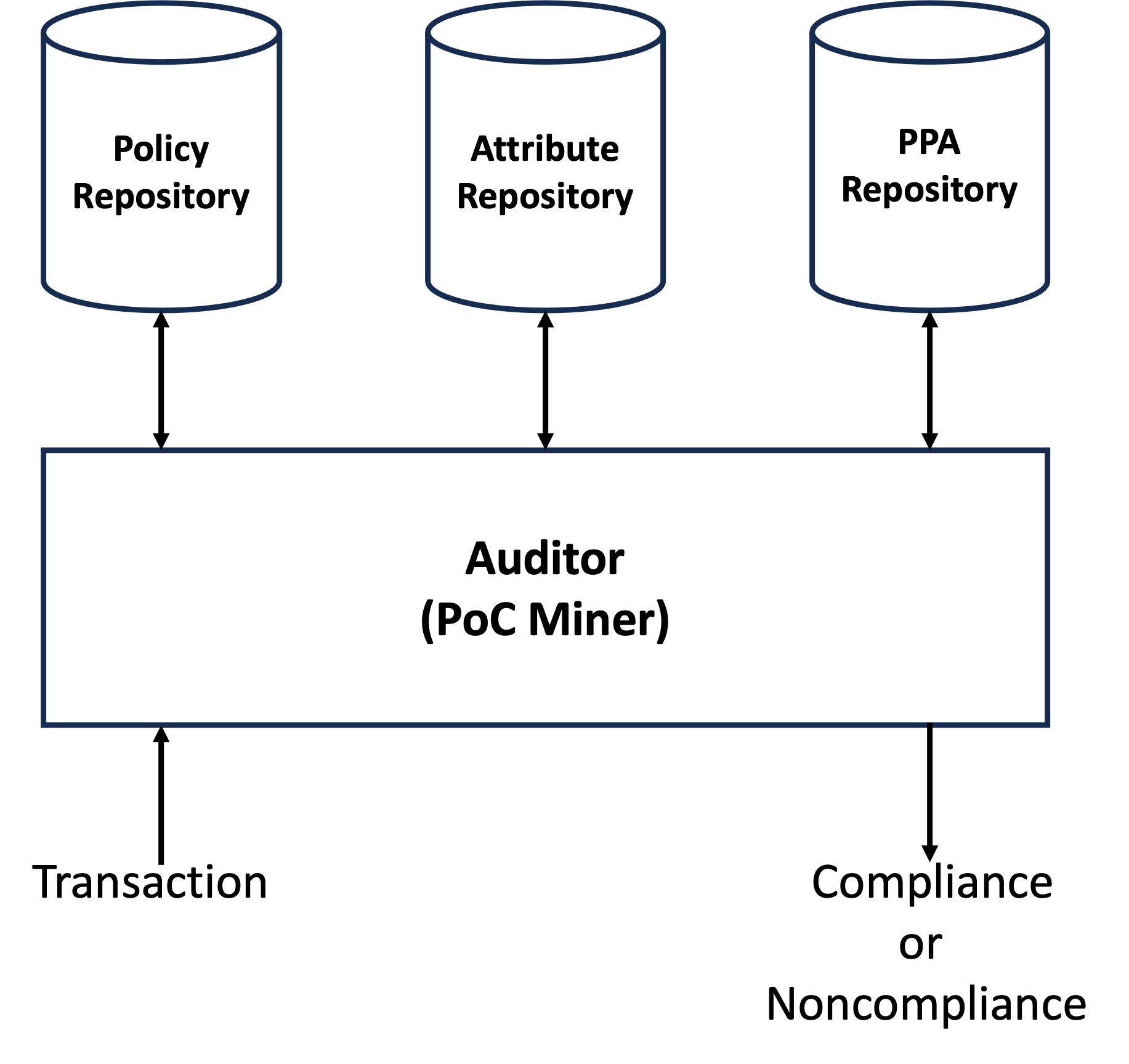}
    \caption{Auditor Nodes Compliance Checking.} \label{fig:compliance-checking}
    \end{center}
\end{figure}

Subjects are users who can perform certain operations on healthcare resources when certain conditions are satisfied. They must be authorized to perform any operations. Unauthorized users should not access protected health information, which can cause security and privacy violations. Operations represent the system actions that authorized users can perform on the objects. Examples of operations are read, write, and update. Objects are protected healthcare information. Any operations based on this information must be according to the applicable policies. A timestamp is the block creation time. The time has been given in seconds since 1.1.1970. The proposed framework captures the time when an access happens.  Conditions that must be satisfied by the user to perform operations on the protected objects. Based on the consensus mechanism, the auditor nodes determine the compliance status. Fig. \ref{fig:compliance-checking} shows the compliance determination process. The status can be \textit{compliance, noncompliance, or not determined.} Selected auditor nodes perform the compliance-checking operations.

\subsection{Proof of Compliance Network Structure}
\begin{figure}[htb]
    \begin{center}
    \includegraphics[scale=0.38]{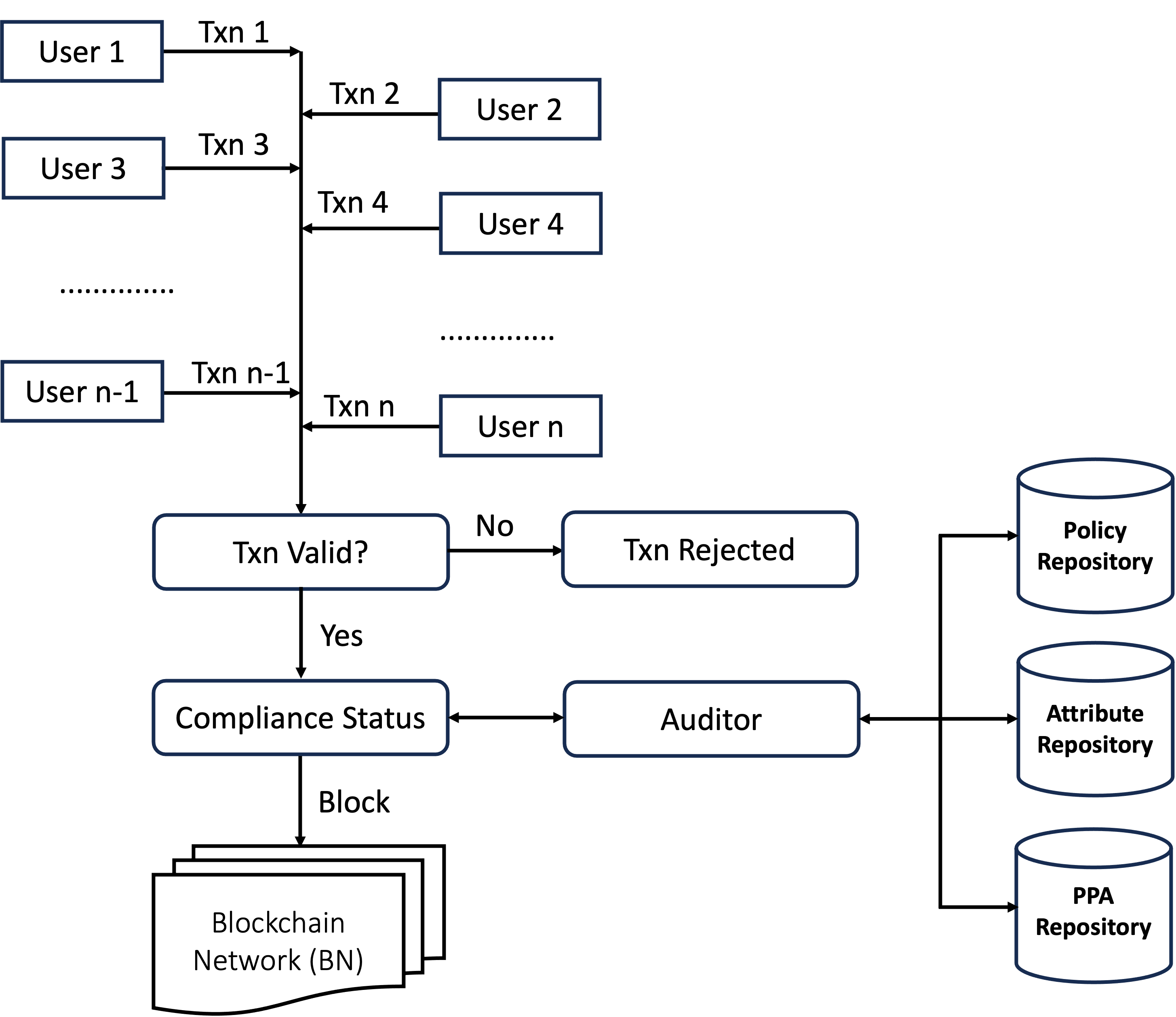}
    \caption{PoC Blockchain Network Structure.} \label{fig:poc-blockchain-network-structure}
    \end{center}
\end{figure}

Fig. \ref{fig:poc-blockchain-network-structure} shows the blockchain structure for the proposed consensus mechanism. All users' access requests are authorized and recorded as audit trails in provenance. Then the proof of compliance consensus mechanism performs compliance checking and adds compliance status as \textit{compliance} or \textit{non-compliance}. Finally, blocks are created and added to the blockchain network.

\subsection{Proof of Compliance (PoC) Participant Nodes}  
Multiple nodes participate in the network to perform different roles. In the following, they are discussed along with their corresponding activities in the network.

    \textbf{Client Node:} This node submits all transactions to the network for confirmation, auditing, and commit. This node is also responsible for reading blocks from the network with the required credentials.

    \textbf{Order Node:} It performs all transaction ordering services submitted by the client node. Transaction ordering can be done in various ways: first come, first serve, priority-based, transaction size, etc. Once transactions are ordered, it transfers them to the validator/endorser node for verification.

    \textbf{Validator/Endorser Node:} Chaincodes, smart contracts, or business logic are executed by this node to validate the transactions. 

    \textbf{Auditor Node:} This node is responsible for checking the compliance requirements for regulations and other applicable bodies.

    \textbf{Committer Node:} It writes the transactions verified by the validator/endorser and auditor nodes to the blockchain ledger. After writing the transactions to the ledger, no entity can modify the blocks or transactions.

\subsection{Proof of Compliance (PoC) Consensus Mechanism}
Algorithm \ref{alg:proof-of-compliance} contains the step-by-step activities of the proof of compliance mechanism. There are four major groups of operations \textit{(i) signature verification and order}, \textit{(ii) transaction validation}, \textit{(iii) policy compliance verification}, and \textit{(iv) ledger modification.} In signature verification and order,  the order nodes perform users' signature verification for submitted transactions. After verification, only valid transactions are ordered for the next step. If a user submits a transaction using a private key but includes another user's public key or ID, then the submitted transaction is invalid. The validator/endorser node executes chaincodes or smart contracts to verify the transaction output. Then, the auditor nodes perform transaction compliance status checking. After checking, a status, compliance or non-compliance, is added at the end of the transaction, as shown in Fig. \ref{fig:transaction-structure}. Other components in the submitted transaction should not be modified. Finally, after adding the compliance status to the transaction, it is permanently added to the blockchain ledger. Before adding, the responsible nodes must verify that the submitted transaction elements are the same.

The PoC reports do not support final regulatory compliance certification. However, it is possible if one or more multiple audit nodes are deployed and maintained in the consensus mechanism by the corresponding regulatory agency, government agency, or compliance authority. Also, it is important to satisfy certain requirements by the proposed framework. The regulatory authority authorities and corresponding responsible offices must provide those requirements. Currently, there is no proposal or requirements for compliance status checking. 

\RestyleAlgo{ruled}
\SetKwComment{Comment}{/* }{ */}

\begin{algorithm}[htb]
\scriptsize
    \SetKwInOut{KwData}{Input}
    \SetKwInOut{KwResult}{Output}
    \SetKw{KwBy}{by}
\DontPrintSemicolon

\caption{Proof of Compliance (PoC) Consensus}\label{alg:proof-of-compliance}

\KwData{ (i) list of transactions $(Txns)$ and (ii) set of policy $Plcy$}
\KwResult{\;
        (i) list of accepted/rejected transactions $(Txns)$ \;
        (ii) list of transactions that are policy compliance}
\;
\textbf{Initialization} \vfill 
$\mathbb{N}_{Order}$ order nodes\;
$\mathbb{N}_{Validator}$ validator/endorser nodes\;
$\mathbb{N}_{Audit}$ audit nodes\;
$\mathbb{N}_{Committer}$ committer nodes\;

\;
 \textbf{Signature Verification and Order} \vfill
         
        $Txn_{Valid} = []$  \textcolor{blue}{\Comment*[r]{accepted transaction list }}\;
        $Txn_{Invalid} = []$  \textcolor{blue}{\Comment*[r]{rejected transaction list }}\;

                        \For{$i \gets Txns_{Start}$ \KwTo $Txns_{End}$ \KwBy $1$}{
                                \eIf{$\zeta (PK_i,Tnx_i) == Signed_{Tnx_i}$}{
                                        $Txn_{Valid} \gets Txn_{Valid} + Txn_i$\;
                                }{
                                        $Txn_{Invalid} \gets Txn_{Invalid} + Txn_i$\;
                                }
                        }
\;

 \textbf{Transaction Validation} \vfill
        $Txn_{Accepted} = []$  \textcolor{blue}{\Comment*[r]{accepted transaction list }}\;
        $Txn_{Rejected} = []$  \textcolor{blue}{\Comment*[r]{rejected transaction list }}\;

                        \For{$i \gets {Txn_{Valid}}_{Start}$ \KwTo ${Txn_{Valid}}_{End}$ \KwBy $1$}{
                                \eIf{$\zeta (PK_i,Tnx_i) == Signed_{Tnx_i}$}{
                                        $Txn_{Accepted} \gets Txn_{Accepted} + {Txn_{Valid}}_i$\;
                                }{
                                        $Txn_{Rejected} \gets Txn_{Rejected} + {Txn_{Valid}}_i$\;
                                }
                        }
 \;
 \textbf{Policy Compliance Verification} \vfill

        $Txn_{Compliance} = []$  \textcolor{blue}{\Comment*[r]{compliance transactions }}\;
        $Txn_{NonCompliance} = []$  \textcolor{blue}{\Comment*[r]{noncompliance transactions}}\;

        \For{$i \gets {Txn_{Accepted}}_{Start}$ \KwTo ${Txn_{Accepted}}_{End}$ \KwBy $1$}{
                                \eIf{$\zeta (PK_i,Tnx_i) == Signed_{Tnx_i}$}{
                                        $Txn_{Compliance} \gets Txn_{Compliance} + {Txn_{Accepted}}_i$\;
                                }{
                                        $Txn_{NonCompliance} \gets Txn_{NonCompliance} + {Txn_{Accepted}}_i$\;
                                }
                        }

 \;
 \textbf{Ledger Modification} \vfill

        $Txn_{Compliance} = []$  \textcolor{blue}{\Comment*[r]{compliance checked final transactions }}\;
        $Txn_{NonCompliance} = []$  \textcolor{blue}{\Comment*[r]{noncompliance transactions}}\;

        \For{$i \gets {Txn_{Accepted}}_{Start}$ \KwTo ${Txn_{Accepted}}_{End}$ \KwBy $1$}{
                                \eIf{$\zeta (PK_i,Tnx_i) == Signed_{Tnx_i}$}{
                                        $Txn_{Compliance} \gets Txn_{Compliance} + {Txn_{Accepted}}_i$\;
                                }{
                                        $Txn_{NonCompliance} \gets Txn_{NonCompliance} + {Txn_{Accepted}}_i$\;
                                }
                        } 
\end{algorithm}

\subsection{Healthcare Data Security and Privacy}
The auditor nodes verify the compliance status of healthcare workers' access activities. For this, they need to access the deployed and applicable policies with the corresponding audit trails. In this situation, the audit nodes do not access any protected health information directly or indirectly. However, activity logs can contain some sensitive information about health professionals as well as patients. For example, it is possible to learn about a patient's ongoing treatment from the audit trails. If a specialist doctor accesses a patient's health data, then it can be concluded that the doctor is treating the patient. In this proposed $PoC$ consensus mechanism, the auditor nodes act as secured, trusted, and blind entities. They don't reveal data to other unauthorized users. They don't analyze data to learn about providers and patients. They do not modify or tamper with data or $PoC$ decisions. They do not store data intentionally. They analyze data only to make compliance status decisions. The complete protocols for them to act as secured, trusted, and blind entities are not addressed in this paper currently. They are our future research directions.

\subsection{ Blockchain Data as Court Evidence}
In June 2018, a landmark court judgment affirmed for the first time that electronic data stored on a blockchain could be considered valid electronic evidence. Subsequently, in a separate case in 2019, the court extended this recognition to not only the authenticity and integrity of electronic evidence stored on a blockchain but also to evidence generated by the blockchain itself \cite{wu2020electronic}. The Blockchain Technology Act, House Bill 3575, passed by the House of Representatives on May 29, 2019, and enacted in January 2020, governs the utilization of blockchain technology in transactions and legal proceedings. The Act sets forth regulations, imposes limitations, and provides definitions for terms such as blockchain and electronic record. According to the Act, the use of a blockchain to create, store, or verify a smart contract, record, or signature does not undermine its legal effect or enforceability. Moreover, when the law mandates written records, electronic evidence recorded on a blockchain is deemed sufficient. Similarly, for situations requiring a signature, an electronically recorded signature on a blockchain or blockchain evidence confirming a person's intent to provide a signature is considered satisfactory \cite{blockgeeksInteractionBetween}. 

Hence, the implementation of our prototype, utilizing blockchain for smart contract storage, is following US law. The hospital can submit complaints about policy violations against offenders, supported by evidence from our blockchain framework.

\section{Achieving Provenance and Compliance} \label{sec:achieving-provenance-compliance}

\begin{figure*}[tb]
    \centering
    \includegraphics[scale=0.42]{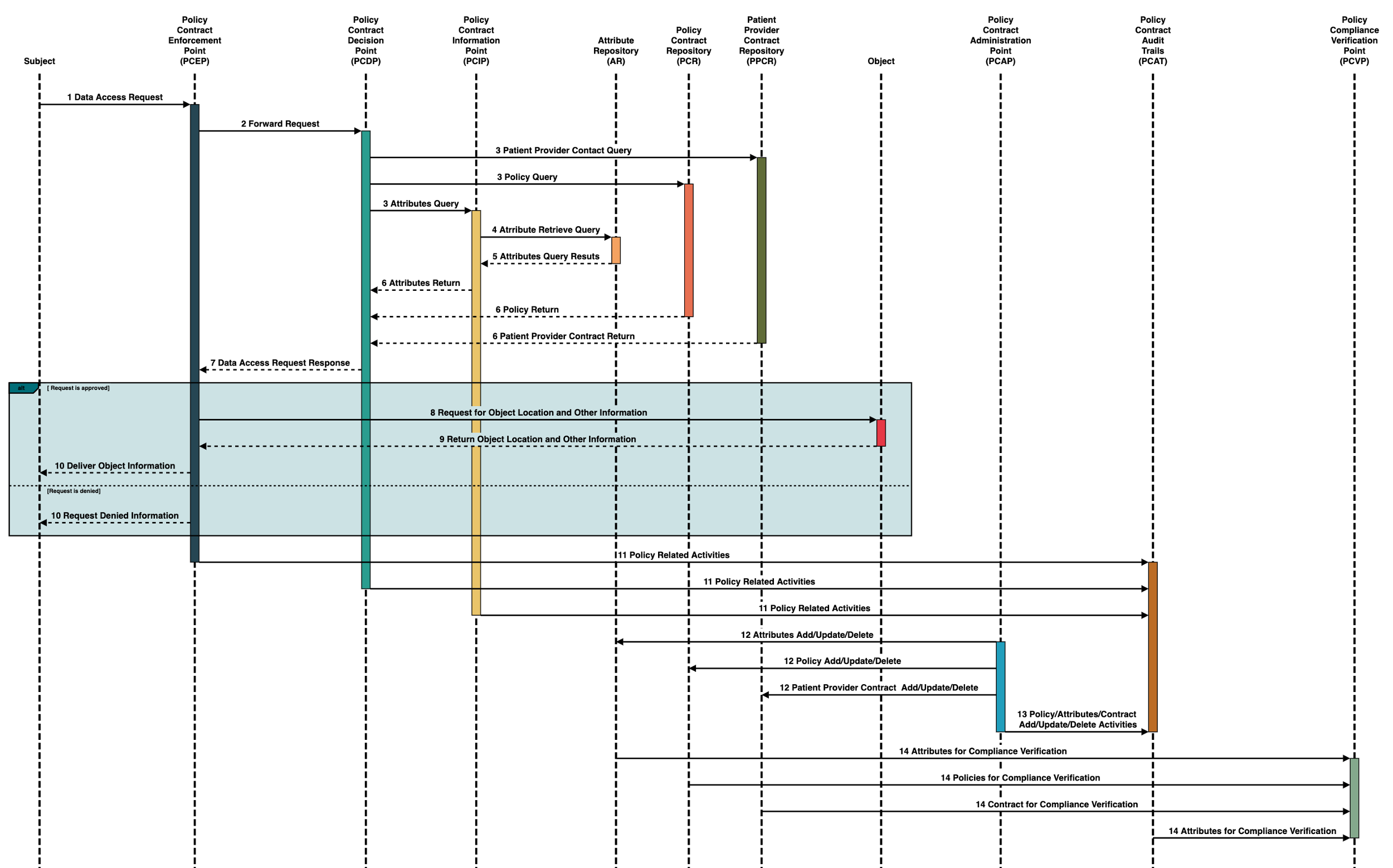}
    \vspace{-1.5em}
    \caption{Policy Enforcement, Provenance, and Compliance Sequence Diagram.}
    \label{fig:compliance-sequence-diagrram}
\end{figure*}

The data access control process for provenance and compliance is summarized in Figure \ref{fig:compliance-sequence-diagrram}, and the steps are as follows:

\textbf{Step 1 and 10:} The subject puts the request to \textit{PCEP} in \textit{Step 1} with the required credentials and attributes. The subject receives the object in \textit{Step 10} if an access request is granted. Otherwise, the subject receives the access denial decision.

\textbf{Step 2 and 7:} \textit{PCEP} forwards the subject access request in \textit{Step 2} with the credentials and attributes received from the subject to \textit{PCDP} to make the decision. In \textit{Step 7}, \textit{PCEP} receives the access decision made by \textit{PCDP}.

\textbf{Step 3, 4, 5, and 6:} In \textit{Step 3}, \textit{PCDP} requests \textit{PCIP} to retrieve the attributes of the subject and object and environmental conditions related to the access request. \textit{PCDP} also asks the policy repository to find the applicable policies. It also requests that the PPAR retrieve the PPAs between the patient and the provider for the access request. In \textit{Step 5}, \textit{PCIP} retrieves the attributes from the attribute repository. \textit{PCDP} receives responses in \textit{Step 6}.

\textbf{Step 8 and 9:} If access is granted, then \textit{PCEP} gets the object in these steps.

\textbf{Step 12:} \textit{PCAP} updates the attribute, policy, and patient-provider agreement repository.

\textbf{Step 11 and 13:} In \textit{Step 11}, \textit{PCEP}, \textit{PCDP}, and \textit{PCIP} record the activities as audit trails to \textit{PCAT}. When \textit{PCAP} updates repositories, its activities are recorded in \textit{Step 13}.

\textbf{Step 14:} \textit{PCVP} gets all required information from repositories to certify policy compliance.

\section{Implementation and Experimental Evaluation}  \label{sec:implementation}

To study the challenges in using blockchain smart contracts for policy compliance provenance as outlined above, we implement a proof-of-concept on the Ethereum blockchain. The proofs-of-concept have two parts: one implements traditional access control policies for a hypothetical healthcare provider and the other implements policies involved in the contract between a patient and the provider.  We implement both policies as contracts per the algorithms shown in Section. 

We chose the Ethereum blockchain protocol test network, Goerli, since it is the first blockchain platform with smart contract features. We use the Python-based Brownie blockchain development framework to build, deploy, test, and do other things with smart contracts. We use Solidity as a programming language for the proposed approach to writing smart contracts. 

We send transactions through the Ethereum test network, Goerli, using the Metamask wallet. Goerli faucet ether is spent for smart contract deployment and transaction submission. In smart contracts, all communication occurs through function calls and event triggers, which the Ethereum protocol handles automatically.

\subsection{Sample Policies}
To make sure that the treatment team members have access to patients' health, we include some sample policies. Other users must not access the data.

\textbf{Policy 1 (P1):} \textit{In an emergency, treatment team member doctors can read and write, and other members can read patients’ health records. Patient and emergency contact personnel must be notified about the emergency access.}

\textbf{Policy 2 (P2):} \textit{In an emergency, emergency contact personnel can read patients’ health records. Patient and emergency contact personnel must be notified about the emergency contact personnel access.}

\textbf{Policy 3 (P3):} \textit{Treatment team member doctors can read and write all medical data for the assigned patient.}

\textbf{Policy 4 (P4):} \textit{Treatment team members, nurses, and support staff can only read particular health records for the assigned patient.}

\textbf{Policy 5 (P5):} \textit{Lab technicians can read and write test results prescribed by doctors. They can not read or write any other results or healthcare records.}

\textbf{Policy 6 (P6):} \textit{Assigned billing officers can read and write patient personal information and billing information. They can not read or write healthcare records.}

\textbf{Policy 7 (P7):} \textit{Assigned insurance company employees can read only medical test names and doctor's information. They can not read or write any medical records, including visit notes.  Insurance coverage claim payment information must be shared with the patient.}

\textbf{Policy 8 (P8):} \textit{A patient can read all his or her medical records, including visit notes, prescriptions, medical test results, and insurance claim information. But he/she can not write any medical records. He/she can write personal, billing, emergency contact, and check-in information.}

\textbf{Policy 9 (P9):} \textit{Audit logs integrity must be protected once they are recorded to prevent log tampering. It means no one can alter the log data.}

\textbf{Policy 10 (P10):} \textit{Healthcare Providers cannot access patients' data if they do not have HIPAA/HITECH/GDPR training from 6/12 months.}

\subsection{Patient Treatment Team Members}
Treatment team members for a patient include doctors, nurses, support staff, lab technicians, billing officers, insurance agents, and the patient's emergency contact person. As the treatment period for a patient, we include everything from treatment to insurance coverage and billing.  For this study, we consider one member from each category, like the doctor, nurse, and others. However, there might be multiple team members in each category. Table \ref{table:patient-treatment-team-member} shows the patient-treatment team members and their responsibilities during and after treatment. Team members are randomly selected if they are available. An emergency contact person is added by the patient and not randomly selected for the treatment team. Algorithm \ref{alg:patient-treatment-team-creation} shows the steps of treatment team creation for a patient.

\begin{table}[tb]
\caption{Patient Treatment Team Member}
\label{table:patient-treatment-team-member}
\vspace{-1em}
\resizebox{\columnwidth}{!}{
\begin{tabular}{|c|l|l|} 
\hline
\rowcolor{lightgray} \textbf{SN} & \textbf{Treatment Team Member}    &  \textbf{Treatment Team Member Responsibilities} \\
\hline
 1 & Doctor (DOC) & Viewing patient's healthcare data\\
 \hline 
 2 & Nurse (NRS) & Creating new patient's healthcare data\\
 \hline 
 3 & Support Staff (STF) & Correcting erroneous or appending patient's healthcare data\\
 \hline 
 4 & Billing Officer (BLO) & Viewing patient's healthcare data\\
 \hline 
 5 & Radiology Lab Tech (RLT) & Creating new patient's healthcare data\\
 \hline 
 6 & Pathology Lab Tech (PLT) & Correcting erroneous or appending patient's healthcare data\\
 \hline 
 7 & Emergency Contact (EMC) & Viewing patient's healthcare data\\
 \hline 
 8 & Pharmacist (PHR) & Creating new patient's healthcare data\\
 \hline 
 9 & Insurance Agent (INA) & Correcting erroneous or appending patient's healthcare data\\
 \hline 
\end{tabular}
}
\end{table}

\RestyleAlgo{ruled}
\SetKwComment{Comment}{/* }{ */}

\begin{algorithm}[tb]
    \footnotesize
    \SetKwInOut{KwData}{Input}
    \SetKwInOut{KwResult}{Result}
    \DontPrintSemicolon

\caption{Patient Treatment Team (PTT) Creation}\label{alg:patient-treatment-team-creation}
\KwData{(i) $DOC$, (ii) $NRS$, (iii) $STF$, (iv) $BLO$, (v) $RLT$, (vi) $PLT$, (vii) $EMC$, (viii) $PHR$, (ix) $INA$}
                \textcolor{blue}{\Comment*[r]{$\mathbb{R}_{PTT}$: secured PTT repository, $BN_{SC}$: blockchain network smart contract}}
\KwResult{A formal treatment team}

\textbf{PTT Member Initialization} \vfill 
  $PTT_i \gets  \{ DOC_i, NRS_i, STF_i, BLO_i, RLT_i, PLT_i, EMC_i, PHR_i, INA_i\}$ for patient identity $i$ \vfill
    \quad \textit{(i)} $DOC \gets \{ DOC_1, DOC_2, DOC_3,.......,DOC_D \}$\; \vfill
    \quad \textit{(ii)} $NRS \gets \{ NRS_1, NRS_2, NRS_3,.......,NRS_N \}$\; \vfill
    \quad \textit{(iii)} $STF \gets \{ STF_1, STF_2, STF_3,.......,STF_S \}$\; \vfill
    \quad \textit{(iv)} $BLO \gets \{ 	BLO_1, BLO_2, BLO_3,.......,BLO_B \}$\; \vfill
    \quad \textit{(v)} $RLT \gets \{ RLT_1, RLT_2, RLT_3,.......,RLT_R \}$\; \vfill
    \quad \textit{(vi)} $PLT \gets \{ PLT_1, PLT_2, PLT_3,.......,PLT_P \}$\; \vfill
    \quad \textit{(vii)} $EMC \gets \{ EMC_1, EMC_2, EMC_3,.......,EMC_E \}$\; \vfill
    \quad \textit{(viii)} $PHR \gets \{ PHR_1, PHR_2, PHR_3, ......., PHR_H \}$\; \vfill
    \quad \textit{(ix)} $INA \gets \{ INA_1, INA_2, INA_3, ......., INA_I \}$\; \vfill

 \textbf{PTT Finalization} \vfill 
  \eIf{$PTT_i$ is complete}{
        \textcolor{blue}{\Comment*[r]{complete: presence of all members}}
            \eIf{$(\mathbb{R}_{PTT} + PTT_i)$ contains no conflicts}{    
                   \textit{(i)} do $\mathbb{R}_{PTT} \gets (\mathbb{R}_{PTT} + PTT_i)$\;
                   \textit{(ii)} add $\mathbb{ID}_{PTT_i}$ to patient profile, $\mathbb{P}_i$\;
                   \textit{(iii)} call $\mathbb{BN}_{SC} (\mathbb{ID}_{PTT_i}, \mathbb{H}_{PTT_i})$\; 
                        \textcolor{blue}{ \Comment*[r]{later PTT integrity verification}}
                   
                   \textbf{\textit{Return: }} Success ($PTT_i$ added to $\mathbb{R}_{PTT}$)\;
            }{
                \textbf{\textit{Error: }} $(\mathbb{R}_{PTT} + PTT_i)$ contains conflicts\;
                            \textcolor{blue}{ \Comment*[r]{ $PTT_i$ revision required to add}}
            }
   }{
        \textbf{\textit{Error: }}$PTT_i$ cannot be created (incomplete PTT)\;  
 }
\end{algorithm}


\subsection{Patient Health Records}
The term "Electronic Health Record" (EHR) refers to an electronic version of a patient's medical history that is kept on file by the healthcare provider over time. It may include all the administrative and clinical information pertinent to the patient's care under a specific provider, such as demographics, progress notes, issues, medications, vital signs, previous medical history, immunizations, laboratory information, and radiology reports. For this study, we only consider ten (10) types of health records for each patient. Table \ref{table:patient-records} shows the health record ID, name, description, and potential creators.

\begin{table*}[tb]
\centering
\caption{Patient Health Records}
\label{table:patient-records}
\vspace{-1em}
\scriptsize
\begin{tabular}{|c| l| l| l|} 
\hline
\rowcolor{lightgray}  \textbf{Record ID} &  \textbf{Record Name} & \textbf{Record Description}  & \textbf{Record Creator}    \\
\hline
 HR1001 & Demographic Info & Patient's information& Patient, Support Staff\\ 
 \hline 
 HR1002 &  Previous Medical History & Old medical records from another hospital & Patient, Support Staff\\
\hline
 HR1003 & Immunizations  & Immunization records that are administered over time & Patient, Pathology Lab Technician \\
\hline
 HR1004& Allergies & Various allergies sources, triggering condition, remediation & Patient, Support Staff, Pathology Lab Technician\\
\hline
 HR1005 & Visit Notes & Contains physiological data, disease description, advice, follow-up, visit details & Doctor, Nurse \\
\hline
 HR1006& Medications and Prescription & Prescribed medications including name, dosage, etc. & Doctor \\
\hline
 HR1007 & Pathology Lab Works & Blood work & Pathology Lab Technician \\
\hline
 HR1008 & Radiology Lab Works & Imaging and Radiology Lab results & Radiology Lab Technician \\
\hline
 HR1009 & Billing and Insurance & Bank account and insurance policy Information & Patient, Support Staff, Billing Officer \\
 \hline
 HR1010 & Payer Transactions & Bills of doctor visit, lab works, and medications & Billing Officers, Insurance Agent \\
\hline
\end{tabular}
\end{table*}


\subsection{Health Record Operations}
Many operations are executed in the healthcare industry to perform required operations. Some common operations are \textit{view/read}, \textit{add/write}, \textit{update/modify}, \textit{share}, \textit{delete}, etc. This study considers only three operations: \textit{read}, \textit{write}, and \textit{update}. Users of patients' health records perform these operations. In the \textit{read} operation, users can only view or read healthcare records or resources if the request is valid and complies with all applicable policies. The state of the data is not changed in this operation, ensuring data integrity. However, it can compromise confidentiality and privacy if the access requested is granted without appropriate credentials. On the other hand, the \textit{write} operation changes the state of the records or data. If proper policy enforcement is not ensured, it breaks the integrity of the data. The \textit{update} operation also changes health data to correct errors or adds new data at the end of the old data. Same as \textit{write}, it changes data integrity. Table \ref{table:health-record-operations} contains the details of the operations with ID, name, and functionalities. In our future Communications, other operations like share and delete will be added to the framework.

\begin{figure}[tb]
\centering
\includegraphics[width=\linewidth]{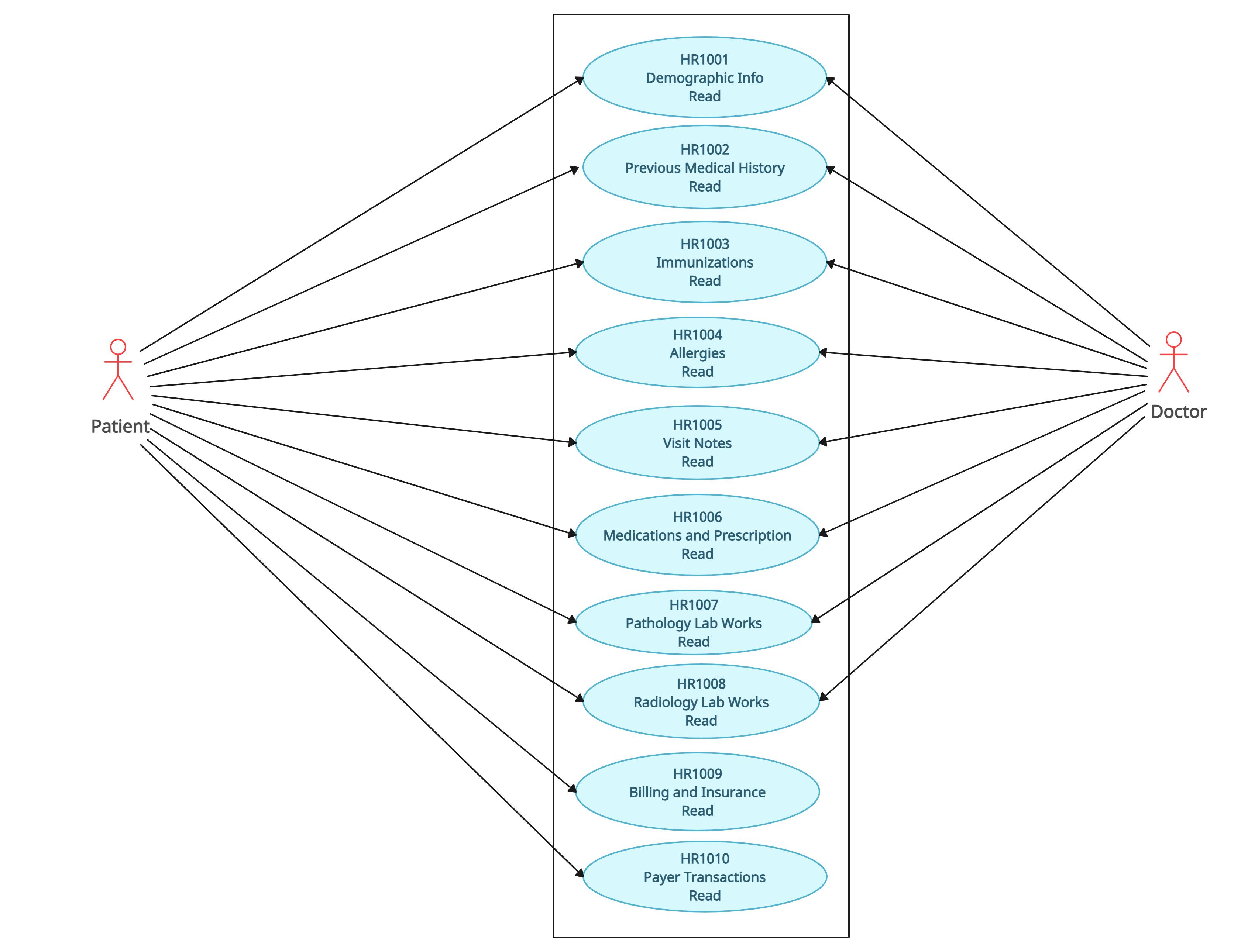}
\vspace{-2em}
\caption{Patient and Doctor Use Cases for Read Operation.}
\label{fig:patient-doctor-use-cases-read}
\end{figure}

\begin{figure}[tb]
\centering
\includegraphics[width=\linewidth]{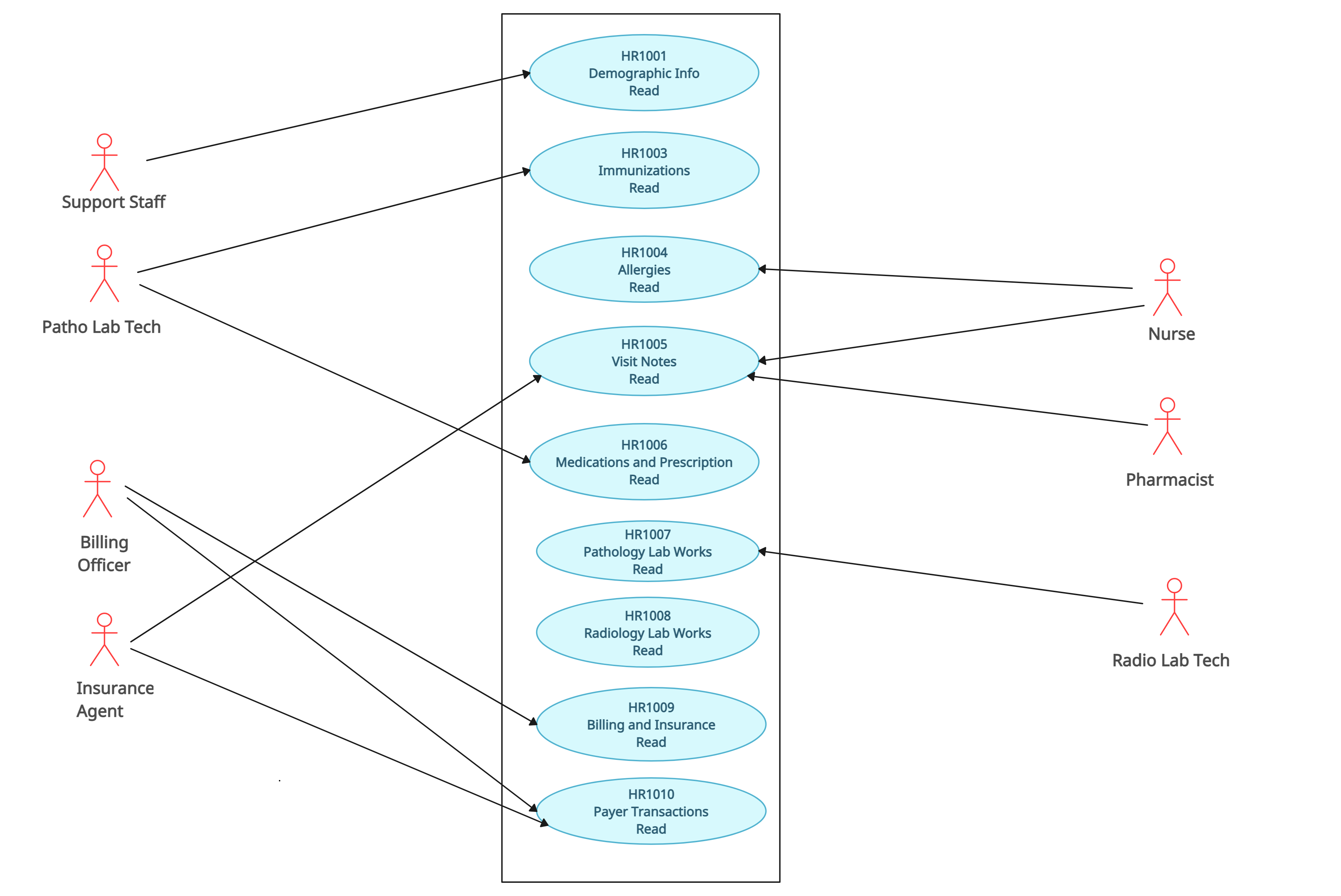}
\vspace{-2em}
\caption{Users Use Cases for Read Operation.}
\label{fig:user-use-cases-read}
\end{figure}

\begin{figure}[tb]
\centering
\includegraphics[width=\linewidth]{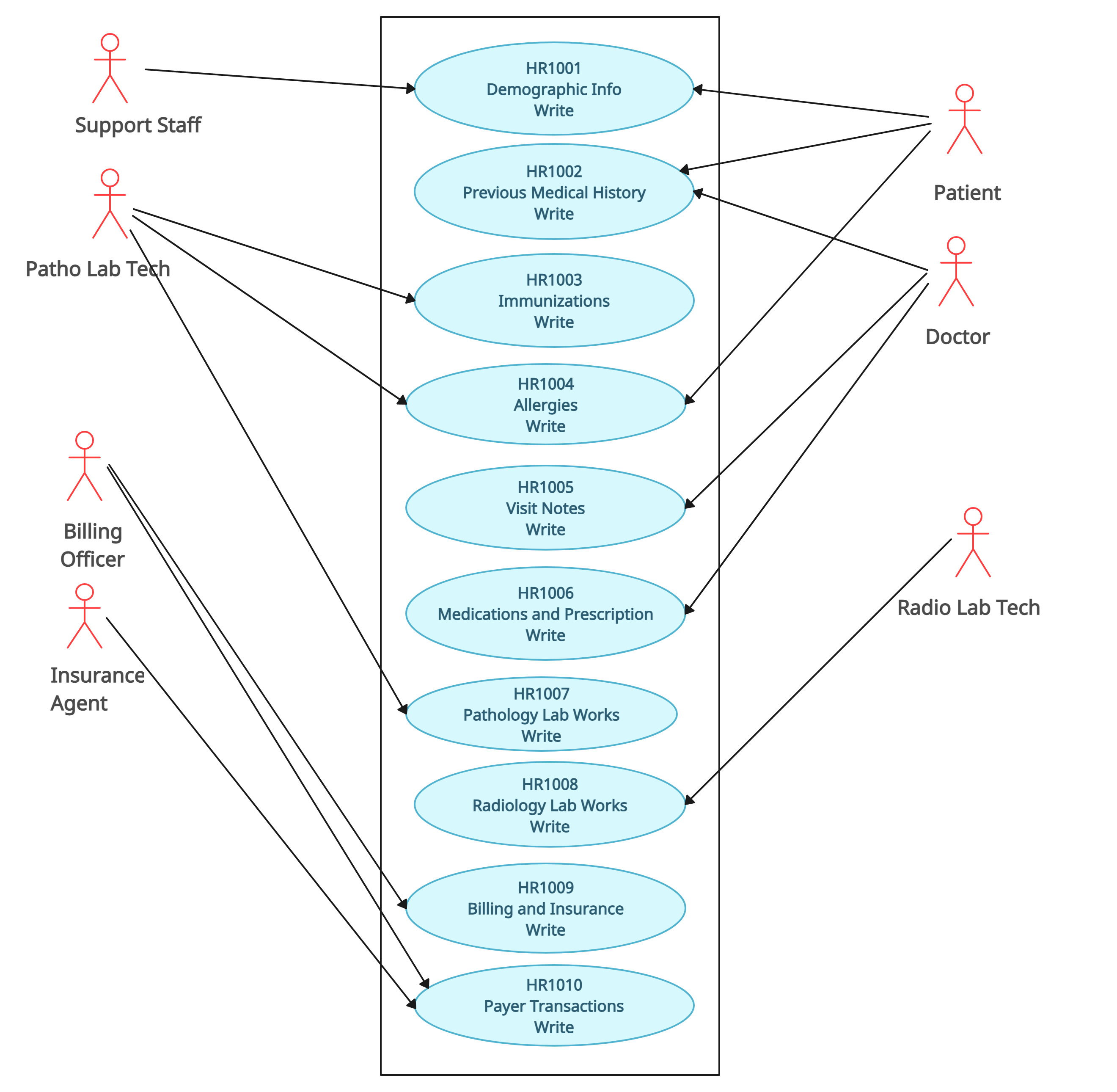}
\vspace{-2em}
\caption{Users Use Cases for Write Operations.}
\label{fig:users-use-cases-write}
\end{figure}

\begin{figure}[tb]
\centering
\includegraphics[width=\linewidth]{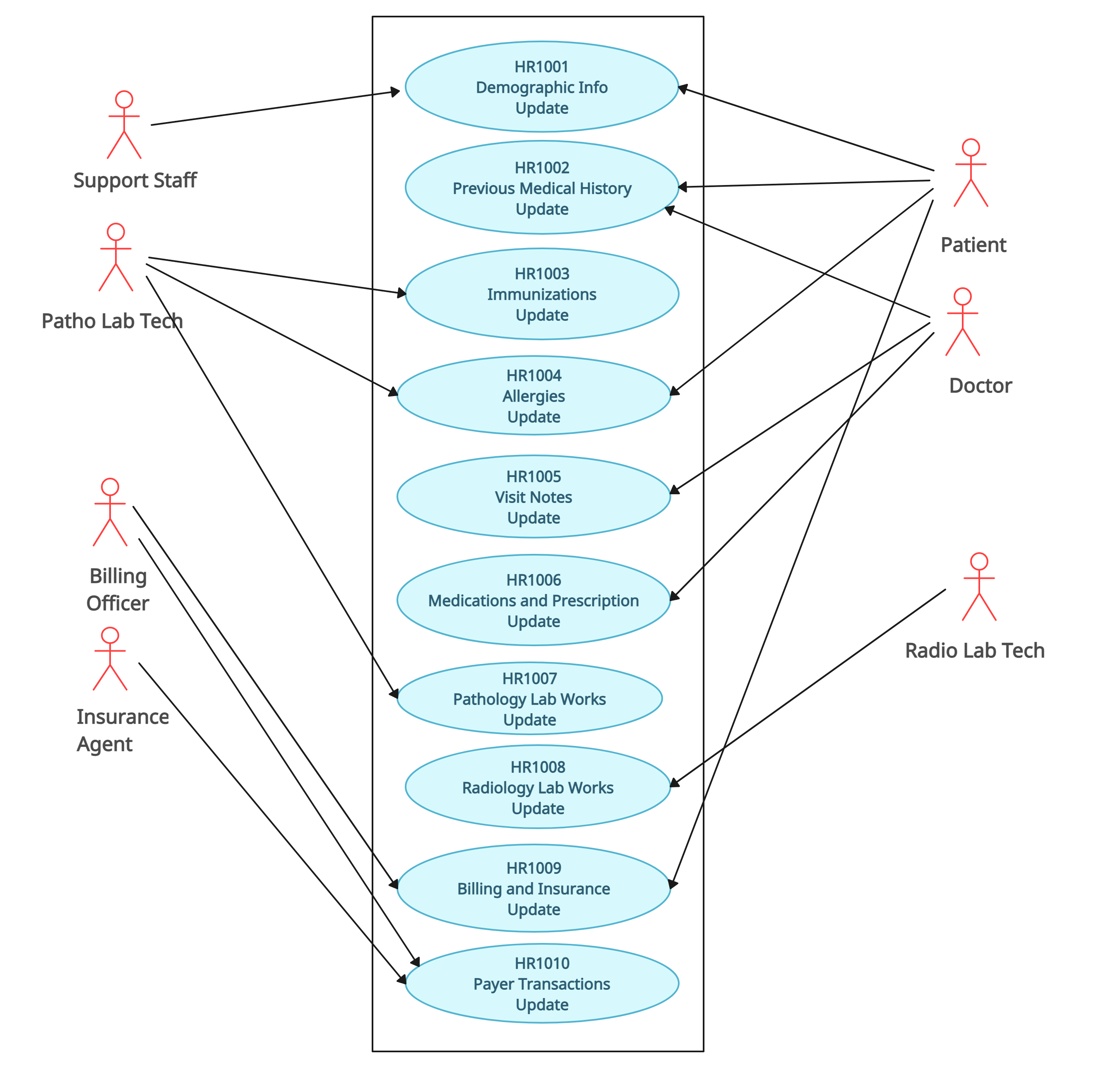}
\vspace{-2em}
\caption{Users Use Cases for Update Operations.}
\label{fig:users-use-cases-update}
\end{figure}

HIPAA privacy policy mandates that users should only access a patient's protected health information if they have permission or consent from the patient. However, there are some cases where a patient's consent is not needed such as the billing process, and anonymous data sharing with research organizations and marketing agents. HIPAA security policy mandates to put the right access control mechanism to protect patient's protected health information. We keep the proposed framework to focus on treatment team-based data access. All users should not perform all operations on all health records. In this work, users have operations permission based on their job role or functionalities. Also, health data nature and sensitivity are considered to give the permissions. Table \ref{table:user-oriented-health-records-operations} shows the user-oriented operations. Fig. \ref{fig:patient-doctor-use-cases-read}, \ref{fig:user-use-cases-read}, \ref{fig:users-use-cases-write}, and \ref{fig:users-use-cases-update} show the use cases diagrams for users and their given operation permissions.

\begin{table}[tb]
\caption{Health Record Operations}
\label{table:health-record-operations}
\vspace{-1em}
\resizebox{\columnwidth}{!}{
\begin{tabular}{|c|l|l|} 
\hline
\rowcolor{lightgray} \textbf{ID} & \textbf{Operation}    & \textbf{Functionalities} \\
\hline
 R & Read & Viewing patient's healthcare data\\
 \hline 
 W & Write & Creating new patient's healthcare data\\
 \hline 
 U & Update & Correcting erroneous or appending patient's healthcare data\\
 \hline 

\end{tabular}
}
\end{table}

\begin{table*}[tb]
\centering
\caption{User Oriented Health Records Operations}
\label{table:user-oriented-health-records-operations}
\vspace{-1em}
\scriptsize
\begin{tabular}{|c | l | l |  l | l |} 
\hline
\rowcolor{lightgray} \textbf{Record ID} &  \textbf{Record Name}  & \textbf{Read} & \textbf{Write} & \textbf{Update}    \\
\hline
HR1001 & Demographic Info  & Patient, Doctor, Support Staff, EC &  Patient, Support Staff &  Patient, Support Staff \\
 \hline 
 HR1002 &  Previous Medical History  & Doctor, Patient & Patient, Doctor & Patient, Doctor \\
\hline
 HR1003 & Immunizations   & Doctor, Patient, Patho Lab Tech& Patho Lab Tech & Patho Lab Tech\\
\hline
 HR1004& Allergies  & Doctor, Patient, Nurse & Patient, Patho Lab Tech & Patient, Patho Lab Tech\\
\hline
 HR1005 & Visit Notes   & Doctor, Nurse, Patient, EC & Doctor & Doctor\\
\hline
 HR1006& Medications and Prescription  & Doctor, Patient, Nurse, Pharmacist, Insurance Agent, EC & Doctor& Doctor\\
\hline
 HR1007 & Pathology Lab Works  & Patho Lab Tech, Doctor, Patient, EC & Patho Lab Tech & Patho Lab Tech\\
\hline
 HR1008 & Radiology Lab Works & Radio Lab Tech, Doctor, Patient, EC &  Radio Lab Tech & Radio Lab Tech\\
\hline
 HR1009 & Billing and Insurance & Patient, Billing Officer, Insurance Agent & Billing Officer, Patient & Billing Officer, Patient\\
 \hline
 HR1010 & Payer Transactions  & Patient, Billing Officer, Insurance Agent & Billing Officer, Insurance Agent  &  Billing Officer, Insurance Agent \\
\hline
\end{tabular}
\end{table*}

\subsection{Sample Users Profile}
 Users perform different operations according to their job responsibilities and applicable policy set. If they follow the policy, then the operations are in policy compliance; otherwise, they are violations.  Users are grouped into major five categories: \textit{((i) patient}, \textit{(ii) emergency contact}, \textit{(iii) providers}, \textit{(iv) pharmacist}, and \textit{(v) insurance agents}. Other users such as business associates, clearing houses, etc are not considered in this report. They would be included in the future extension works. Table \ref{table:patient-profiles} shows ten (10) patients' information with six (6) attributes: patient ID, name, date of birth, gender, phone, and email. For every patient, an emergency contact person is included who is the contact point if there is any emergency. Table \ref{table:emergemncy-contact-profiles} contains ten (10) emergency contact information with ID, name, date of birth, phone, email, patient ID, and relationship with the patient.

Table \ref{table:provider-profiles} tabulates ten (10) providers' information with provider ID, name, date of birth, title, phone, and email. For this study, we consider doctors, nurses, support staff, radiology lab technicians, pathology lab technicians, and billing officers are the main providers. Pharmacists' and insurance agents' information is given in Table \ref{table:pharmacist-profiles} and \ref{table:insurance-agent-profiles}. They also access protected health information for processing prescriptions and insurance claims. We keep them as part of the patient treatment team members. They must be authorized to access any patient data. Patient's consent must be given to them. Otherwise, the proposed framework doesn’t allow them to access data.

\begin{table}[tb]
\caption{Patient Profiles}
\label{table:patient-profiles}
\vspace{-1em}
\resizebox{\columnwidth}{!}{
\begin{tabular}{ |c|l|l|l|l|l|} 
\hline
\rowcolor{lightgray} \textbf{Patient ID} & \textbf{Name}    & \textbf{Date of Birth}  & \textbf{Gender} & \textbf{Phone}  & \textbf{Email}   \\
\hline
 PT1001 & Jordan & DOB11251980 &  Male & +15306524342 & jordam@compliance.com\\
 \hline 
 PT1002 & Simon & DOB10151982 & Male & +15737524481 & simon@compliance.com\\
 \hline 
 PT1003 & Tatum & DOB11051984 & Male & +14324386527 & tatum@compliance.com\\
 \hline 
 PT1004 & Thomas & DOB12201986 &  Male & +16609079598 & thomas@compliance.com\\
 \hline 
 PT1005 & David & DOB05091975 & Male & +15702917315 & david@compliance.com\\
  \hline
 PT1006 & Alexander & DOB01271978 & Male & +16578059479 & alexander@compliance.com\\
  \hline
 PT1007 & Sarah & DOB09281970 & Female & +12058912490 & sarah@compliance.com\\
  \hline
 PT1008 & Ronald & DOB03151979 & Male & +13238648870 & ronald@compliance.com\\
 \hline 
 PT1009 & Rebecca & DOB12251985 &  Female & +14426746222 & rebecca@compliance.com\\
 \hline 
 PT1010 & Emma & DOB10151995 & Female & +16098632161  & emma@compliance.com\\
 \hline
\end{tabular}
}
\end{table}

\begin{table}[tb]
\caption{Emergency Contact Profiles}
\label{table:emergemncy-contact-profiles}
\vspace{-1em}
\resizebox{\columnwidth}{!}{
\begin{tabular}{ |c|l|l|l|l|l|l|} 
\hline
\rowcolor{lightgray} \textbf{Id} & \textbf{Name}    & \textbf{Date of Birth}   & \textbf{Phone} & \textbf{Email} & \textbf{Patient}  & \textbf{Relationship} \\
\hline
 EC1001 & Olivia & DOB11041992 & +15135030830 & olivia@service.info & PT1001 & Sister\\
 \hline
 EC1002 & Isabella & DOB12281972 &  +12246579011 & isabella@service.info & PT1002 & Aunt\\
 \hline
 EC1003 & Amelia & DOB07011999 &   +15166694568 & amelia@service.info & PT1003 & Mother\\
 \hline 
 EC1004 & Alice & DOB12251998 & +12164898791 & alice@service.info & PT1004 & Spouse\\
 \hline 
 EC1005 & Eleanor & DOB09111990 & +14305361691 & eleanor@service.info & PT1005 & Spouse\\
 \hline
 EC1006 & Benjamin & DOB10071980 &  +15772628573 & benjamin@service.info & PT1006 & Friend\\
 \hline
 EC1007 & Theodore & DOB06281992 & +15646892293 & theodore@service.info & PT1007 &  Son\\
 \hline 
 EC1008 & Henry & DOB08061982 & +15416532424 & henry@service.info & PT1008 &  Brother\\
 \hline 
 EC1009 & Arthur & DOB12101994 & +16025371089 & arthur@service.info & PT1009 & Brother\\
 \hline
 EC1010 & Liam & DOB02101987 & +13217057450 & liam@service.info & PT1010 & Cousin\\
 \hline

\end{tabular}
}
\end{table}

\begin{table}[tb]
\caption{Provider Profiles}
\label{table:provider-profiles}
\vspace{-1em}
\resizebox{\columnwidth}{!}{
\begin{tabular}{ |c|l|l|l|l|l|l|} 
\hline
\rowcolor{lightgray} \textbf{Provider Id } & \textbf{Name}    & \textbf{Date of Birth}   & \textbf{Title} & \textbf{Phone} & \textbf{Email}  \\
\hline
 PR1001 & Andrew & DOB11041992 & Doctor &+14016296781 & andrew@hospital.com\\
 \hline
 PR1002 & Sophia & DOB12281972 & Doctor &+16805970987 & sophia@hospital.com\\
 \hline
 PR1003 & Linda & DOB07011999 & Doctor &+12522173068 & linda@hospital.com\\
 \hline 
 PR1004 & Oscar & DOB12251998 & Nurse &+12488860746 & oscar@hospital.com\\
 \hline 
 PR1005 & William & DOB09111990 & Nurse &+18026758468 & william@hospital.com\\
 \hline
 PR1006 & Damien & DOB10071980 & Support Staff &+13194615516 & damien@hospital.com\\
 \hline
 PR1007 & Douglas & DOB06281992 & Support Staff &+18597638767 & douglas@hospital.com\\
 \hline 
 PR1008 & Robert & DOB08061982 & Radiology Tech &+18502557057 & robert@hospital.com\\
 \hline 
 PR1009 & Victoria & DOB12101994 & Pathology Tech &+12837975034 & victoria@hospital.com\\
 \hline
 PR1010 & James & DOB02101987 & Billing Officer &+17712104375 & james@hospital.com\\
 \hline

\end{tabular}
}
\end{table}

\begin{table}[tb]
\caption{Pharmacist Profiles}
\label{table:pharmacist-profiles}
\vspace{-1em}
\resizebox{\columnwidth}{!}{
\begin{tabular}{ |c|l|l|l|l|l|l|} 
\hline
\rowcolor{lightgray}  \textbf{Pharmacist ID} & \textbf{Name}    & \textbf{Date of Birth}   & \textbf{Title} & \textbf{Company} & \textbf{Phone} & \textbf{Email}  \\
\hline
 PHR1001 & Justin & DOB12121984 & Pharmacist & EverGreen Pharmacy & +12564014540 & justin@evergreen.phar\\
  \hline
 PHR1002 & Madison & DOB15071977 & Pharm Technician & BlueSky Pharmacy &+15134414566 & madison@bluesky.drug\\
  \hline
 \end{tabular}
}
\end{table}

\begin{table}[tb]
\caption{Insurance Agent Profiles}
\label{table:insurance-agent-profiles}
\vspace{-1em}
\resizebox{\columnwidth}{!}{
\begin{tabular}{ |c|l|l|l|l|l|l|} 
\hline
\rowcolor{lightgray} \textbf{Agent ID} & \textbf{Name}    & \textbf{Date of Birth }  & \textbf{Title}  & \textbf{Company} & \textbf{Phone} & \textbf{Email}   \\
\hline
 ICA1001 & Jasper & DOB21091955 & Senior Agent & Anthem Health Plan & +13524606592  & jasper@antheplan.org\\
  \hline
 ICA1002 & Hanan & DOB17021985 & Agent & Care Health Insurance & +18189025033 & hanan@care.care\\
 \hline

 \end{tabular}
}
\end{table}


\subsection{Provenance Environment}
For policy provenance, we have opted for a private blockchain infrastructure, specifically utilizing the Ethereum private network deployed through the Go implementation of the Ethereum (geth) client. This decision is to ensure data security and maintain control over policy-provenance activities. This private network uses the Proof of Authority (PoA) consensus algorithm, known as Clique, to mine and confirm the audit trails. Further, when we have proof of compliance algorithm ready, the clique algorithm can be modified to change block structure according to requirements.

\begin{figure}[tb]
\centering
\includegraphics[width=\linewidth]{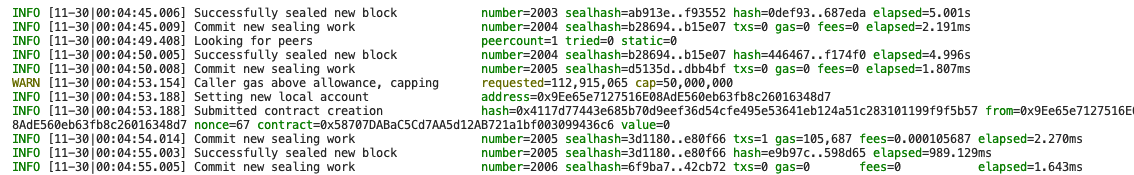}
\vspace{-2em}
\caption{Miner Node Operations.}
\label{fig:miner-node}
\end{figure}

Fig. \ref{fig:miner-node} shows the miner node,  responsible for the end-to-end process of transaction handling. Beginning with the submission of transactions. Once a transaction is submitted, the miner node takes charge of including it in a block and subsequently engages in the mining process to validate the block. Furthermore, the miner node actively publishes this mined data to all other nodes within the network, ensuring a synchronized and updated ledger across the entire system.

\begin{figure}[tb]
\centering
\includegraphics[width=\linewidth]{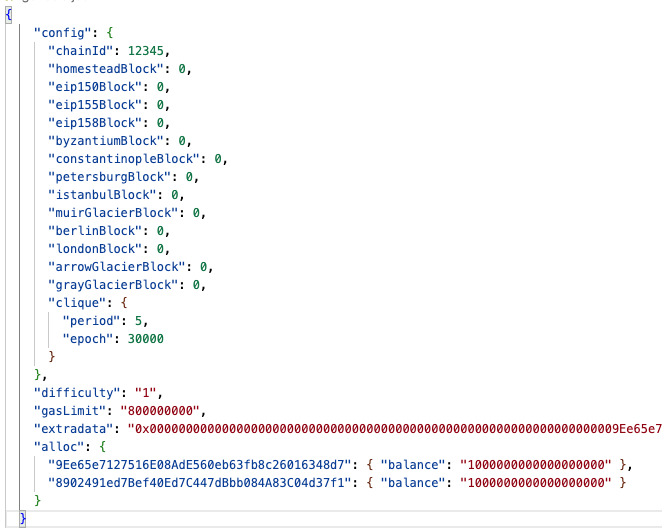}
 \vspace{-2em}
 \caption{Audit Blockchain Genesis Block.} \label{fig:genesis-block}
\end{figure}

Fig. \ref{fig:genesis-block} shows the configuration file for the genesis block of our private network. In the context of blockchain technology, a genesis block is an initial block in the chain, and this JSON file encapsulates crucial information defining the network's foundational parameters. Within this file, the "chainId" attribute is set to "12345," signifying the unique identifier assigned to our private network. The "period" attribute represents a defined period, specifying the block time, which dictates the interval between the creation of consecutive blocks in the blockchain. 

\begin{figure}[tb]
\centering
\includegraphics[width=\linewidth]{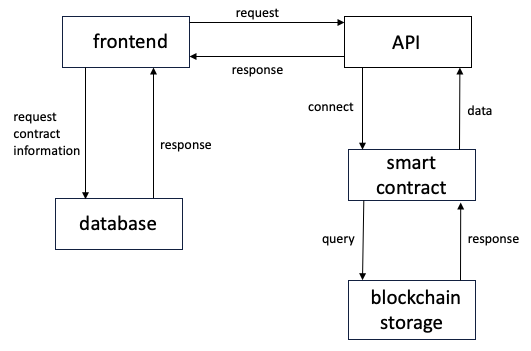}
 \vspace{-2em}
 \caption{Frontend Interaction.} \label{fig:frontend-backend-interaction}
\end{figure}

\begin{figure}[tb]
\centering
\includegraphics[width=\linewidth]{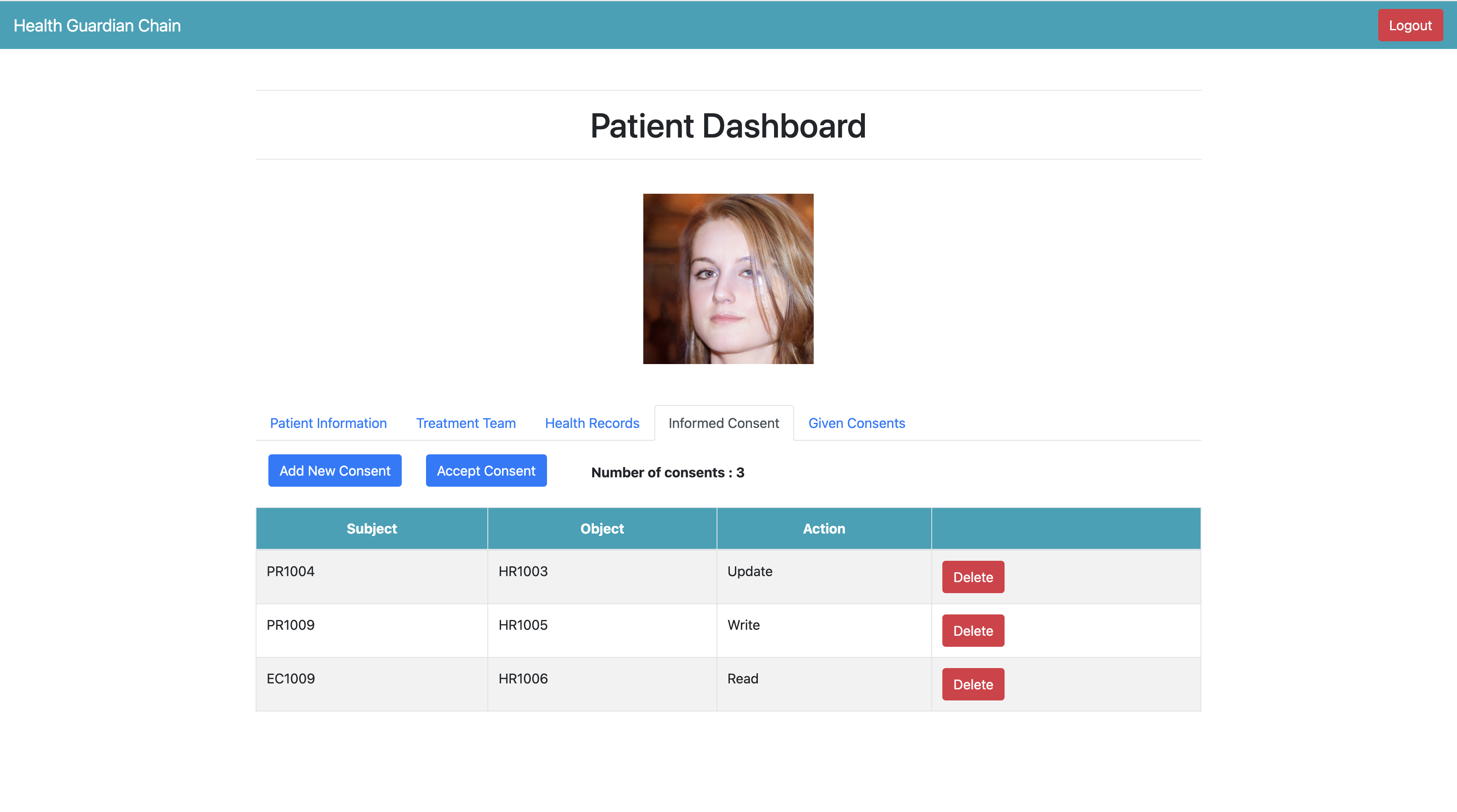}
 \vspace{-2em}
 \caption{User Interface for Patient} \label{fig:user-interface-patient}
\end{figure}

\begin{figure}[tb]
\centering
\includegraphics[width=\linewidth]{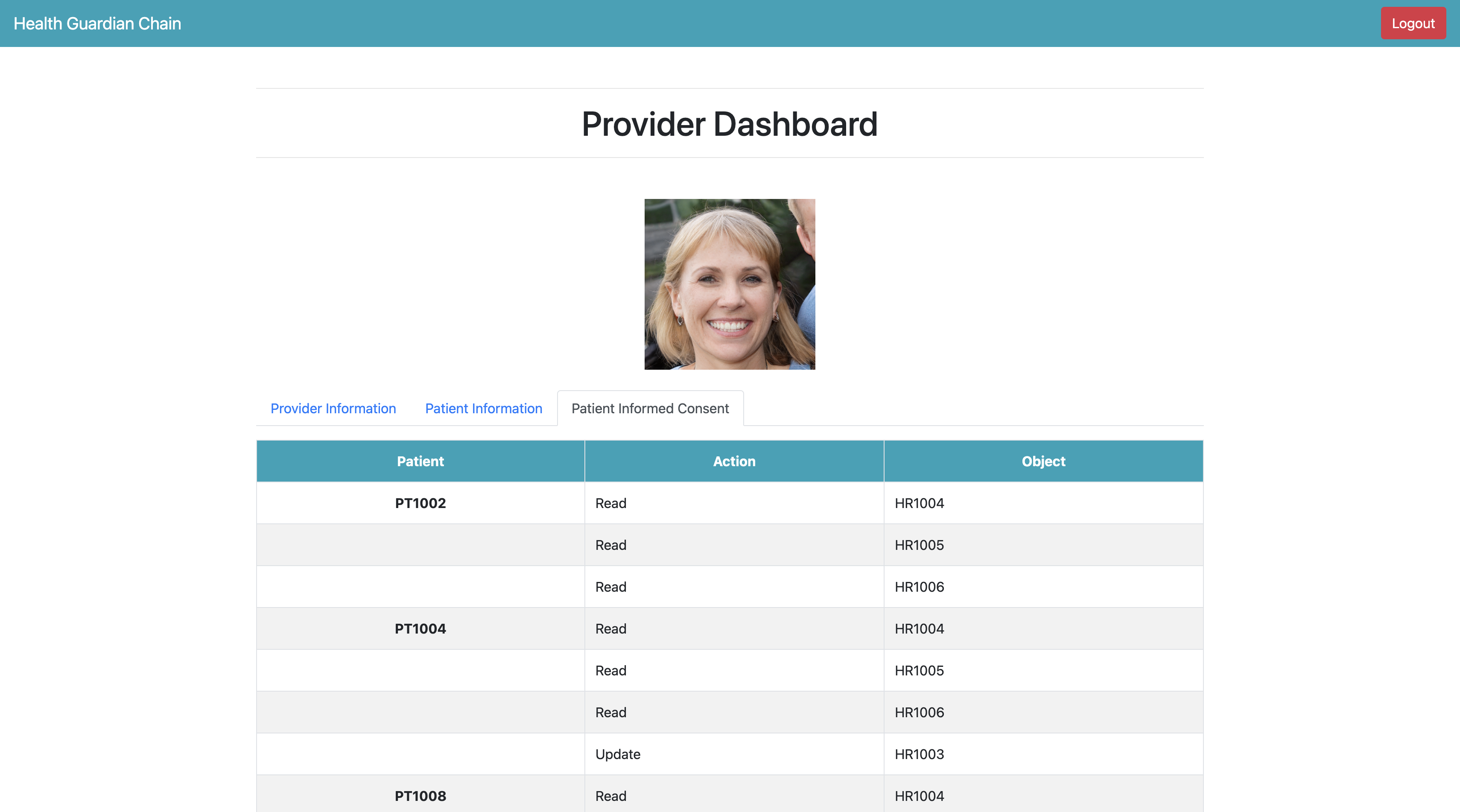}
 \vspace{-2em}
 \caption{User Interface for Provider.} \label{fig:user-interface-provider}
\end{figure}

\begin{figure}[tb]
\centering
\includegraphics[width=\linewidth]{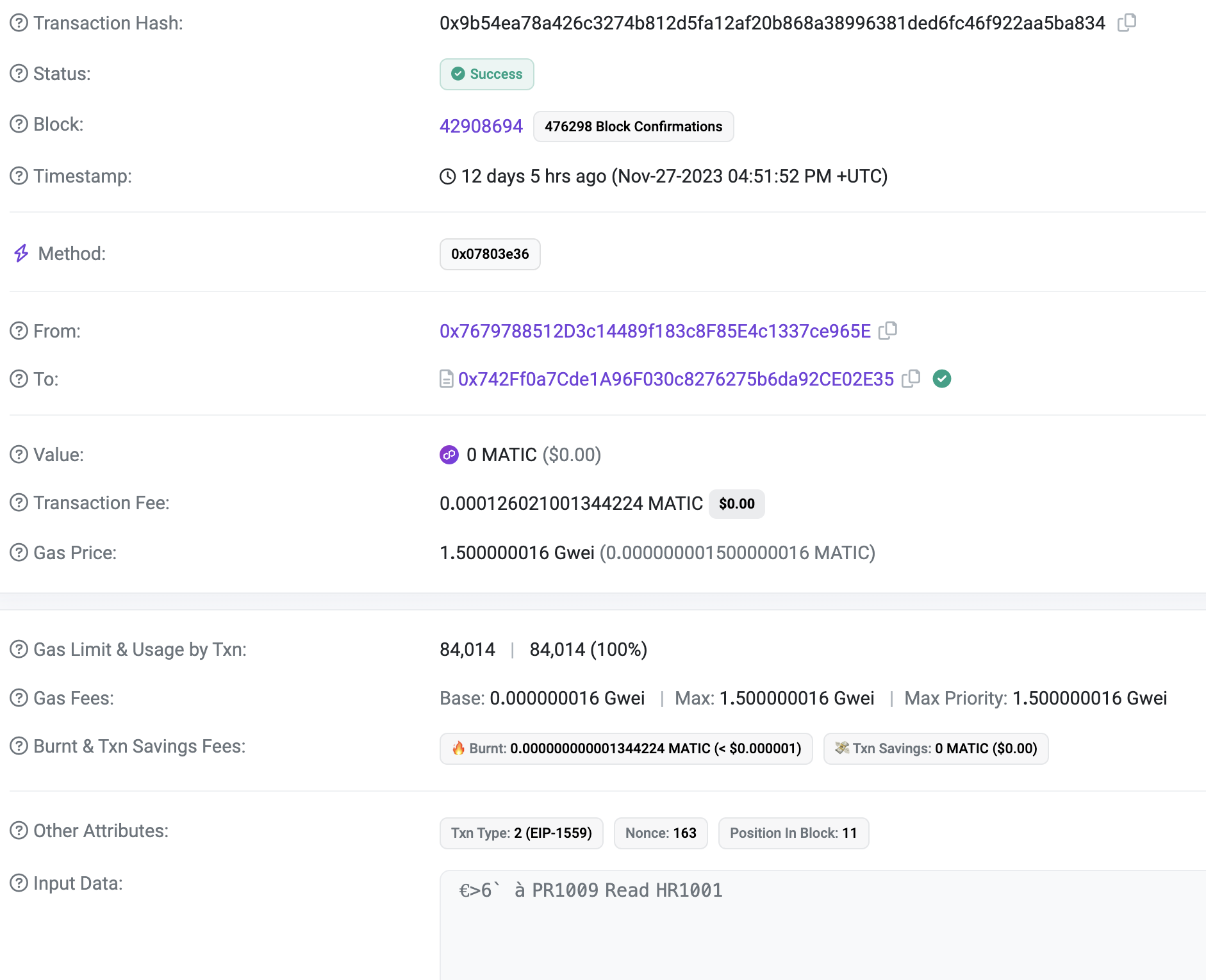}
 \vspace{-2em}
 \caption{Informed Consent Transaction. } \label{fig:informed-consent-transactions}
\end{figure}

\begin{figure}[tb]
\centering
\includegraphics[width=\linewidth]{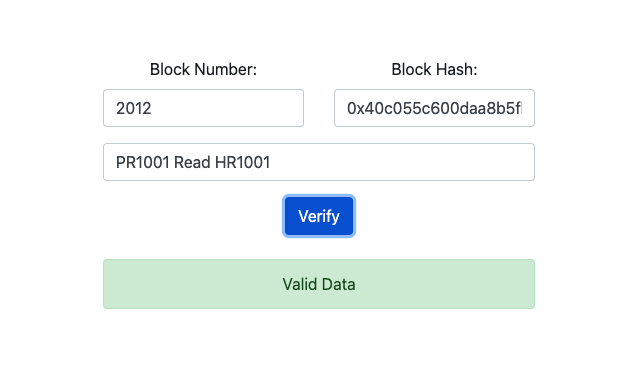}
\vspace{-2em}
\caption{Integrity Verifier.}
\label{fig:verifier}
\end{figure}

Fig. \ref{fig:verifier} shows the system for controlling access to audit trails involves the protection of sensitive user information, executed operations, or health data access. Limited access is granted only to users with designated privileges, following policies.  When a user seeks verification, the Integrity Verifier retrieves the block ID and block hash from the audit blockchain. The verifier then queries the public blockchain for the block hash corresponding to the audit blockchain block ID. If the audit block hash matches the stored hash on the public blockchain, the verifier confirms the data as unaltered. Any mismatch signals potential tampering in the audit blockchain. It's important to note that all blocks in the audit blockchain undergo addition through a consensus mechanism. A single-bit modification invalidates all blocks from the tampered block, indicating any form of alteration. Acting as a blind and trusted entity, the Integrity Verifier, serving as an API or Oracle, ensures data confidentiality, refrains from data modification, and conducts secure verifications for user requests initiated through the user interface.

\section{Conclusion} \label{sec:conclusion}

The paper proposes a blockchain-based approach to address the challenges of healthcare policy compliance. It focuses on enforcing patient-provider agreements (PPAs) and other policies through smart contracts on the blockchain. The use of blockchain technology ensures that policies are enforced and provides an immutable trail for policy compliance. The document highlights the importance of maintaining provenance, which involves recording and preserving the history of policy enforcement activities. Blockchain is used to store audit trails and ensure the integrity of recorded events.

The proposed approach includes a consensus mechanism called Proof of Compliance (PoC) to verify the compliance status of access requests. This mechanism involves a network of distributed and decentralized auditor nodes that perform compliance checking for the given audit trails. The results of the compliance checks determine whether an access request is compliant or non-compliant with the policies.

The paper discusses the components of the proposed approach, including the Patient-Provider Agreement (PPA), informed consent, contract-based access control, policy provenance, and policy compliance. It also provides an overview of the implementation and experimental evaluation of the proposed approach on the Ethereum blockchain.

In conclusion, the paper presents a comprehensive solution for healthcare policy compliance using blockchain technology. The proposed approach ensures that policies are enforced, provenance is maintained, and compliance is verified through the use of smart contracts and distributed auditor nodes. The experimental evaluation demonstrates the effectiveness of the approach in ensuring policy compliance and maintaining the integrity of policy enforcement activities. This research has significant implications for improving the security and privacy of healthcare data and enhancing the trustworthiness of healthcare systems. Future research directions are also discussed to further enhance and extend the proposed compliance frameworks.

\section{Future Directions} \label{sec:future-directions}

We're aiming to put in place a complete step-by-step process for a compliance algorithm and link it up with the system. Additionally, we're looking ahead to expand this system and utilize it to minimize health insurance fraud as part of the proposed compliance framework.

\bibliographystyle{IEEEtran}
\bibliography{main}
\end{document}